\documentclass{./styles/svproc}
\usepackage{graphicx}
\usepackage{caption}
\usepackage{subcaption}
\usepackage{multicol}
\usepackage{footmisc}
\usepackage{amsmath}
\usepackage{listings}

\usepackage{comment} 

\usepackage{url}

\begin{document}
\mainmatter              

\title{Detecting time-fragmented cache attacks against AES using Performance Monitoring Counters}

\titlerunning{Detecting time-fragmented attacks against AES using PMCs}  

\author{Iv\'{a}n Prada, Francisco D. Igual \and Katzalin Olcoz}

\authorrunning{Iv\'{a}n Prada et al.} 

%
%
\tocauthor{Name1 Lastname1, Name2 Lastname2, Name3 Lastname3}

%
%
\institute{Departamento de Arquitectura de Computadores y Autom\'{a}tica, Universidad Complutense de Madrid, Madrid 28040, Spain\\
\email{ivprada@ucm.es}, \email{figual@ucm.es}, \email{katzalin@ucm.es}}

\maketitle

\begin{abstract}
Cache timing attacks use shared caches in multi-core processors as side channels to extract information from victim processes. These attacks are particularly dangerous in cloud infrastructures, in which the deployed countermeasures cause collateral effects in terms of performance loss and increase in energy consumption. We propose to monitor the victim process using an independent monitoring (detector) process, that continuously measures selected Performance Monitoring Counters (PMC) to detect the presence of an attack. Ad-hoc countermeasures can be applied only when such a risky situation arises. In our case, the victim process is the AES encryption algorithm and the attack is performed by means of random encryption requests. We demonstrate that PMCs are a feasible tool to detect the attack and that sampling PMCs at high frequencies is worse than sampling at lower frequencies in terms of detection capabilities, particularly when the attack is fragmented in time to try to be hidden from detection. 

%
%
\end{abstract}


%
\section{Introduction}
In January 2018, Jann Horn from Google Project Zero and a group of researchers led by Paul Kocher independently disclosed three vulnerabilities, named Spectre (variants 1 and 2) and Meltdown \cite{Horn2018}. They discovered that data cache timing could be used to extract information about memory contents using speculative execution.
Since that moment, new variants of these transient execution attacks have been disclosed, such as Foreshadow or NetSpectre, to name just two of them~\cite{Canella2018}.

These attacks exploit speculative and out-of-order execution in high performance microarchitectures together with the fact that in modern multi-core architectures some resources are shared across cores.  Hence, a malicious process which is being executed in one core of the system can extract information from a victim executed in a different core. The most commonly used resource as side-channel to extract information is the shared cache~\cite{Biswas2017}.

This problem is particularly important in cloud environments, where not only multiple users share a multi-core server but also multiple virtual machines can co-reside in the same core due to consolidation in order to save energy. Moreover, the use of simultaneous multithreading techniques, such as Intel's Hyperthreading technology, allow to leverage two or more logical cores per physical core, increasing the degree of resources shared between users.

There has been a proliferation of ad-hoc defenses, mainly microcode and software patches for the operating system and virtual machine monitor. Besides, Intel announced hardware mitigations in its Cascade Lake processors, trying to reduce performance loss due to the countermeasures for some of the attacks~\cite{Kumar}.

However, the impact of countermeasures on performance is still non negligible, and according to~\cite{Canella2018} varies from 0\% to almost 75\%. Thus, in most situations, security comes at the expense of lower performance and higher energy consumption (due to non-consolidating and disabling hyperthreading).

In this paper, we propose a new attack detection tool that is based on the deployment of a process running in the same core that the victim process it protects, and that detects situations in which an attack is being performed. Following this idea, countermeasures are only taken when the risk level justifies the cost.

The contribution of the paper is two-fold:

\begin{itemize}
    \item We implement and describe the attack, and design and implement a detector for it based on Performance Monitoring Counters (PMC) monitoring, evaluating its detection capabilities at different sampling frequencies, showing that high sampling frequencies (100 $\mu$s) are noisier than lower ones.
    \item We show that splitting the attack into small pieces and distributing those pieces in time decreases detection capability in a different way for the different detection sampling frequencies. Only low frequencies such as 10 ms are still able to detect the time-fragmented attack.
\end{itemize}

The rest of the paper is structured as follows: Section~\ref{sec:relatedwork} reviews the most relevant works in the field; Section \ref{sec:background} outlines the main concepts needed for the correct understanding of the attack and detection strategy. Then, the attack implemented, detection using PMC and the time-fragmented attack are presented in Sections~\ref{sec:attack},~\ref{sec:detection} 
and~\ref{sec:time-fragmented}, respectively. Finally, conclusions are presented in Section~\ref{sec:conclusions}.
\section{Related Work} \label{sec:relatedwork}
Detailed surveys on microarchitectural timing attacks in general \cite{Ge2018}, \cite{Biswas2017} and cache timing attacks in particular \cite{Lyu2017} can be found in the literature. \cite{Canella2018} includes a systematic evaluation of transient execution attacks.

Time-driven attacks against the shared and inclusive Last Level Cache (LLC) are mainly based on Flush\&Reload~\cite{Yarom} and their variants. So, both \cite{Bernstein2005} and \cite{Briongos2019} extract the key from the AES T-table based encryption algorithm using improvements over the original attack.

Recently, Performance Monitoring Counters have been used to detect the attack. Chiappeta et al \cite{Chiappetta2016} monitor both the victim and the attacker, while CloudRadar \cite{Bowers2012} monitors all the virtual machines running in the system. CacheShield \cite{Briongos2018a} only monitors the victim process to detect attacks on both AES and RSA algorithms. None of them considered trying to hide the attack by dividing it into small pieces distributed in time.
Our approach is similar to CacheShield~\cite{Briongos2018a} in terms of functionality, but we perform a more detailed study of how the specific timing of the attack affects the detection capability.
\section{Background concepts}
\label{sec:background}

For a correct understanding of the attacks and techniques described hereafter, further details on two 
architectural concepts with direct impact on the attacks are required: {\em cache inclusion policies} and {\em memory de-duplication} as a specific case of shared memory. Then, the basics of the Flush\&Reload attack are outlined.

\subsection{Shared caches and inclusion policies in modern multi-cores}

Modern multi-core processor feature multi-level caches in which levels can be classified as 
{\em shared/private} across cores and hierarchies as {\em inclusive}, {\em non-inclusive}
or {\em exclusive}, 
depending on whether the content of a cache level is present in lower cache levels. Of special 
interest for us is the combination of shared/inclusive cache levels, such as
LLC caches in modern Intel multi-cores; in this scenario, a process executed on a specific 
core can produce side effects on independent processes executed on a different core. This
phenomena can be exploited to perform cache-timing attacks.
Supplementary techniques, such as  Intel's Cache Allocation Technology (CAT~\cite{Nguyen2016}), can be 
leveraged to isolate specific LLC ways in order to boost performance (reducing
contention), but also to mitigate the effects of potential attacks in this 
type of processors and situations.

\subsection{Shared memory and memory de-duplication\label{sec:deduplication}} 

Modern operating systems, such as Linux, make an intensive use of shared memory across
processes to improve memory usage efficiency. Some situations (e.g. parent-child
process hierarchies generated through {\em fork()}) are easily trackable, but sharing
memory pages across independent processes requires ad-hoc sophisticated techniques. This
is a very common scenario in multi-VM deployments sharing the same physical resources,
for example.

Memory de-duplication is a specific technique of shared memory, designed to reduce the memory
footprint in scenarios in which a hypervisor shares memory pages with the same contents
across different virtual machines, but with impact also on non-virtualized environments
comprising random non-related processes.
In the Linux implementation (KSM, {\em Kernel Samepage Merging}), a kernel thread periodically
checks every page in registered memory sections, and calculates a hash of its contents. This
hash is then used to search other pages with identical contents. Upon success, pages
are considered identical and merged, saving memory space. Processes that reference to the original
pages are updated to point to the merged one. Only after a write operation
from one of the VMs (or processes), sharing finishes and the corresponding page is copied
by COW ({\em Copy-on-Write}).

\subsection{Flush\&Reload\label{sec:far}}

The {\em Flush\&Reload} attack was first introduced in~\cite{Yarom} and optimized by later works such as~\cite{Bernstein2005}, among others. It takes advantage of the combination of inclusive shared caches and memory de-duplication. The basics of the attack are as follows: the attacker runs in a core which shares the last level cache with the victim, and manages to share some page with it through memory de-duplication. It can either contain shared data (i.e. the tables used by the AES encryption algorithm) or shared instructions (for the attack against RSA). In the first phase of the attack ({\em Flush}), the attacker evicts the shared blocks from its own private cache, causing the eviction of those data from the shared cache and all the other caches. In the second phase, the victim performs some random work, bringing some of the shared data to the cache again. In the last phase ({\em Reload}), the attacker accesses every shared data, measuring the time it takes and it guesses which data have been used by the victim (cache hits) and which ones were not used (cache misses). From this information, the attacker extracts relevant data, such as the AES key.

\section{Implementation of the AES attack}
\label{sec:attack}

\subsection{Experimental Setup}

The experimental setup was deployed on a dual-socket server featuring two Intel
Xeon Gold 6138
chips with 20-cores each (hyperthreading was disabled), 
running at 2 Ghz. The memory hierarchy comprises 96 Gbytes DDR4 RAM, 
28 Mbytes of unified L3 cache per chip (11-way associative), 
1 Mbyte of unified L2 cache per chip (16-way associative)
and 32 Kbytes of L1 cache per core (8-way associative).  
Cache line is 64 bytes.
L1 TLB comprises 64 entries (4-way associative) with a page size of 4 Kbytes.

From the software perspective, we employed a Debian GNU/Linux distribution with kernel
4.9.51-1 and GCC 6.3.0.
PAPI version 5.5.1.0~\cite{Terpstra2010} built on top of the Linux {\tt perf\_event} subsystem
was employed to extract performance counters information. OpenSSL version 1.1.1.b was used
to implement the cryptographic algorithm, compiled with the {\tt no-asm} flag when using
T-tables.

\subsection{AES algorithm}

In~\cite{Daemen2000}, authors develop the underlying theory of polynomials with coefficients in $GF(2^8)$. This is the base for the extraction of transformation values of a single round. The round transformation lies in four steps for the first rounds ({\sc SubByte}, {\sc ShiftRows}, {\sc MixColumns} and {\sc AddRoundKey}) and three for the last round (all but one, {\sc MixColumns}). The number of rounds will depend on the length of the key; in our case, for 128-bits key, we need $10$ rounds.
As stated by Vicent Rijmen et al. in~\cite{Daemen2000}, the round transformation of AES can be optimized with  $4$ look-up tables that contain the pre-calculated values for each of the potencial inputs $T_i,\ i\in 0\cdots 3$. This way, the encryption round will be simplified to a few XOR operations and takes the form:
\begin{equation}
\label{AES_eq_1}
   S_{i,j}=T_{i}[s^{k}_{i,j}] \oplus RoundKey^{k}_{i,j}
\end{equation}
for the main rounds, and the last round:
\begin{equation}
\label{AES_eq_2}
   S_{i,j}=T_{(i+2)\%4}[s^{10}_{i,j}] \oplus RoundKey^{10}_{i,j}
\end{equation}
with $S^{k}_{i,j}$ the encrypted char,  $s_{i,j}$ the previous state char, $k \in 1\cdots 9$ the $k$-th round and $s_1$ is the original message ($s_{0}$) XOR with $RoundKey^{0}$.

\subsection{Implementation of the attack}

The basis of the attack is simple: using T-Tables optimization to extract the last round key of AES. In Section~\ref{sec:relatedwork}, we exposed previous algorithms for extraction of the AES key. We use the approach of \cite{Briongos2019} to break the OpenSSL 1.1.1.b AES 128 bits implementation (this library has had to be compiled with no-asm flag, so that it uses the T-Tables implementation).
The attack begins by forcing the de-duplication of library pages (see Section~\ref{sec:deduplication}). 
This step is mandatory so that victim and attacker can share pages of the dynamic library, 
hence allowing the observation of memory addresses assigned to AES tables.
In order to obtain the origin of the dynamic library, we proceed by opening the library and performing a memory projection (through {\tt mmap}). Proceeding this way, the KSM daemon will detect a matching in 
the contents of the mapped file and the loaded dynamic library, and will force the de-duplication. 
We have experimentally observed a delay of around 300 encryptions to unleash the de-duplication of pages. 
At that point, the attack can commence.
The start addresses of each table are obtained by decompiling the library and determining the 
offset of each table w.r.t. its start address.

As seen in Section~\ref{sec:relatedwork}, there are different ways to extract the key based on the information
left by the last round of encryption. In this work, we check whether a cache 
line\footnote{A cache line --64 bytes in our target architecture-- can store 16 elements of a table, provided  each element is stored as a 4-byte unsigned integer.} resides in L3 upon completion of the encryption process.

These measurements have been carried out empirically by a {\em Flush\&Reload}
technique (see Section~\ref{sec:far}) for each one of the four tables. In the following, $T_j$ is 
the corresponding line of the observed table; the attack proceeds by first performing a {\em flush} 
operation of different lines of the table, followed by a random encryption request. The response to 
this request is then stored ($S[i]$ stores the encrypted text on the $i$-th encryption), together 
with the information that will be necessary to perform the attack: a matrix $X$ is created and
$X_{ij}$ set to 1 if line $T_j$ was in L3 after completing the $i$-th encryption, 0 otherwise.

Once these data are obtained, we proceed by searching for the most probable characters belonging to the 
last round key, following the pseudo-code depicted in Algorithm~\ref{alg:LRK}, that will return, 
for each position of the last round key, those characters with the lowest probability. Hence, we will select:

\begin{equation}
    LastRoundKey_{i, j} = \min_{t\ \in\ 0,...,num\_encrypt}LRK_{i, j}[t]
\end{equation}

Once the characters of the last round key have been obtained, the last step is just an inversion of the code used by AES to obtain the last round key, and hence the initial key of the server.

\begin{lstlisting}[frame=lines,mathescape=true,basicstyle=\small,numbers=left,stepnumber=1,label={alg:LRK},caption={Pseudo-code to obtain Last Round Key candidates.}]
for t in $0,\cdots,num\_encrypt$
  for i in $0, 1, 2, 3$
    if X[(i+2)%4][t] == 0
      for j in $0, 1, 2, 3$
        for l in $0,\cdots,line\_elems$
          $LRK_{i, j}[S^{t}_{i,j} \oplus T_{(i+2)\%4}[l]]++$
	    end for
	  end for
    end if
  end for
end for
\end{lstlisting}

\section{Attack detection using PMCs}
\label{sec:detection}

Cache timing attacks cause an anomalously high number of L3 misses, due to the flush and reload activity; hence, measuring L3 misses is an straightforward mechanism to detect them. As explained in 
Section~\ref{sec:relatedwork}, there have been some works in this field and most of them use L3 misses.

In addition to L3 cache misses, we chose the total number of load instructions executed by the victim as 
a way to estimate the number of encryptions being performed by the victim, so that the ratio between 
both counters provides a metric that is constant for different levels of load in the victim. 
Thus, our detection metric is the number of L3 cache misses per 1000 load instructions. 

\begin{figure}
\centering

\begin{subfigure}[b]{0.49\textwidth}
        \includegraphics[width=\textwidth]{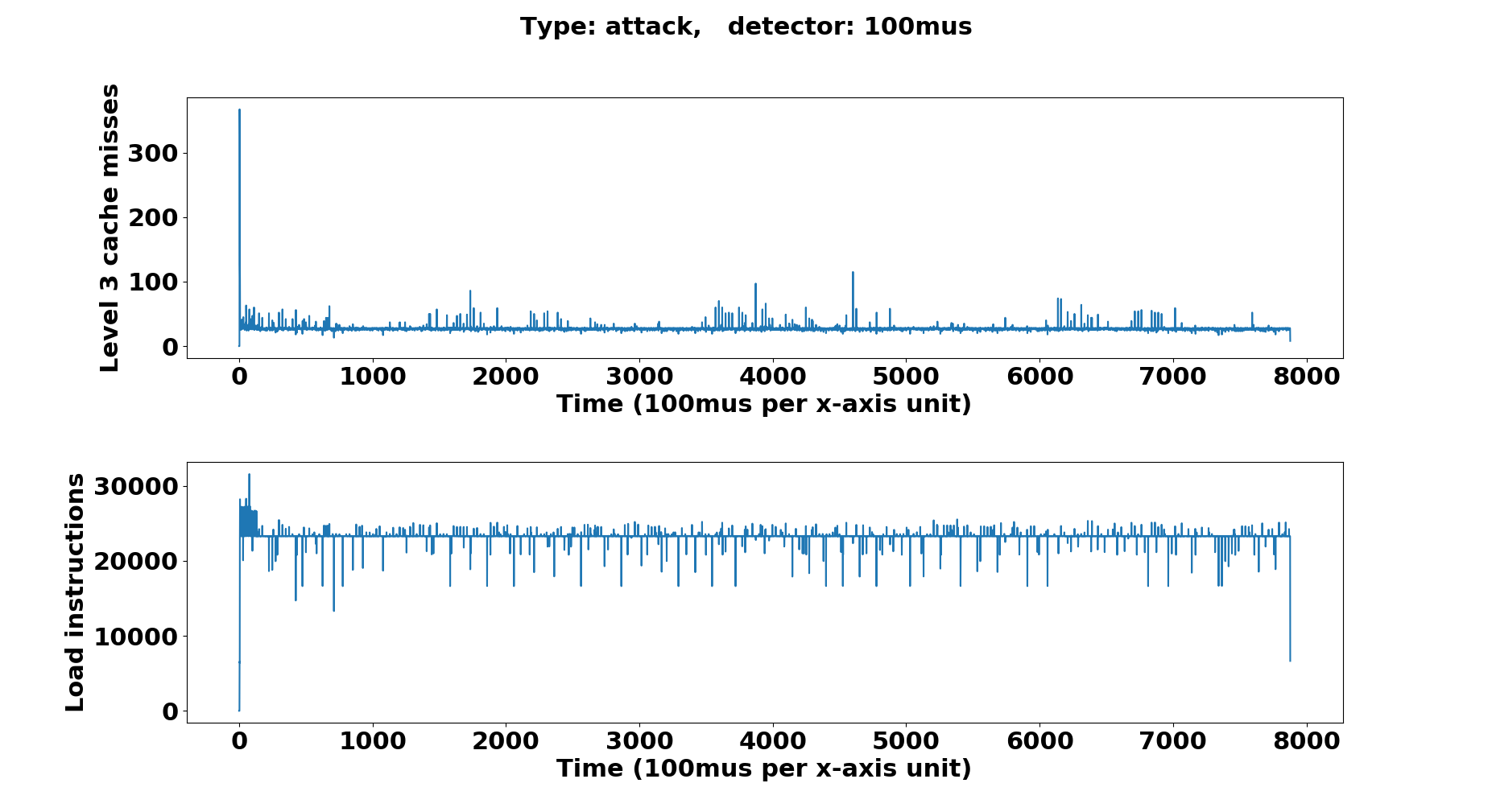}
        \caption{Attack. Detection freq.: 100 $\mu$s\label{fig:subfig11}}
\end{subfigure}
\begin{subfigure}[b]{0.49\textwidth}
        \includegraphics[width=\textwidth]{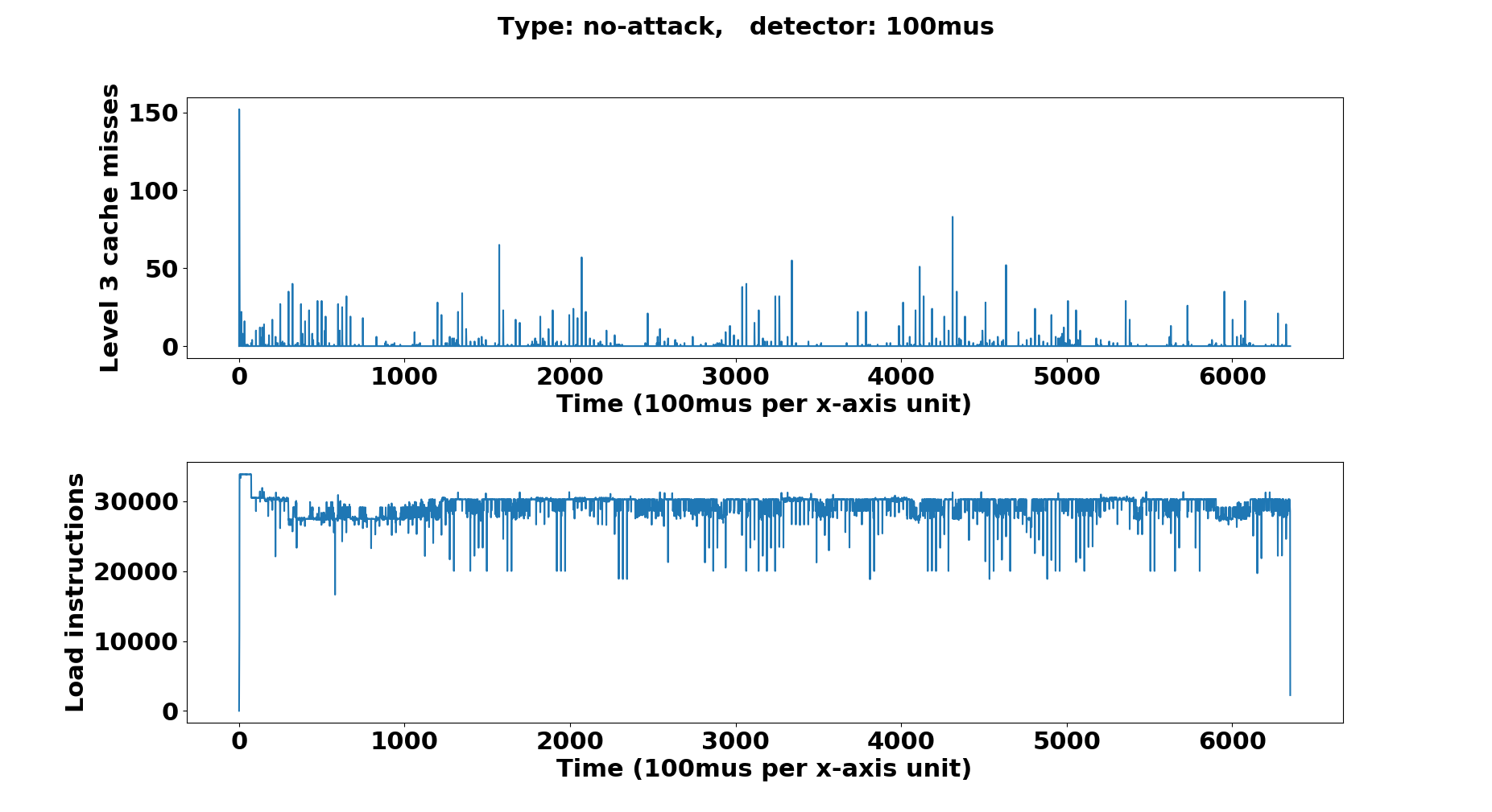}
        \caption{No attack. Detection freq.: 100 $\mu$s}
        \label{fig:subfig12}
\end{subfigure}

\begin{subfigure}[b]{0.49\textwidth}
        \includegraphics[width=\textwidth]{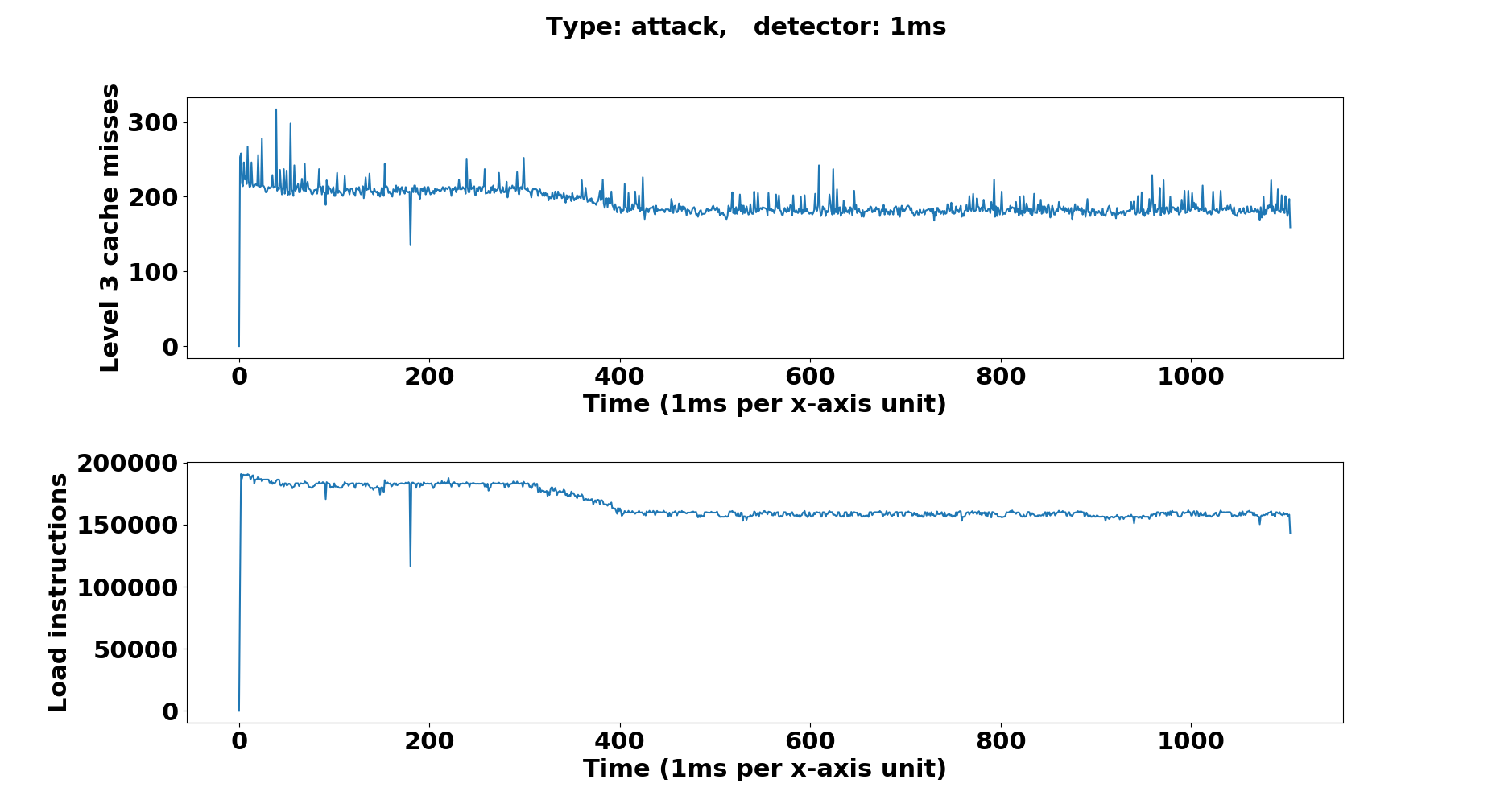}
        \caption{Attack. Detection freq.: 1 ms}
        \label{fig:subfig13}
\end{subfigure}
\begin{subfigure}[b]{0.49\textwidth}
        \includegraphics[width=\textwidth]{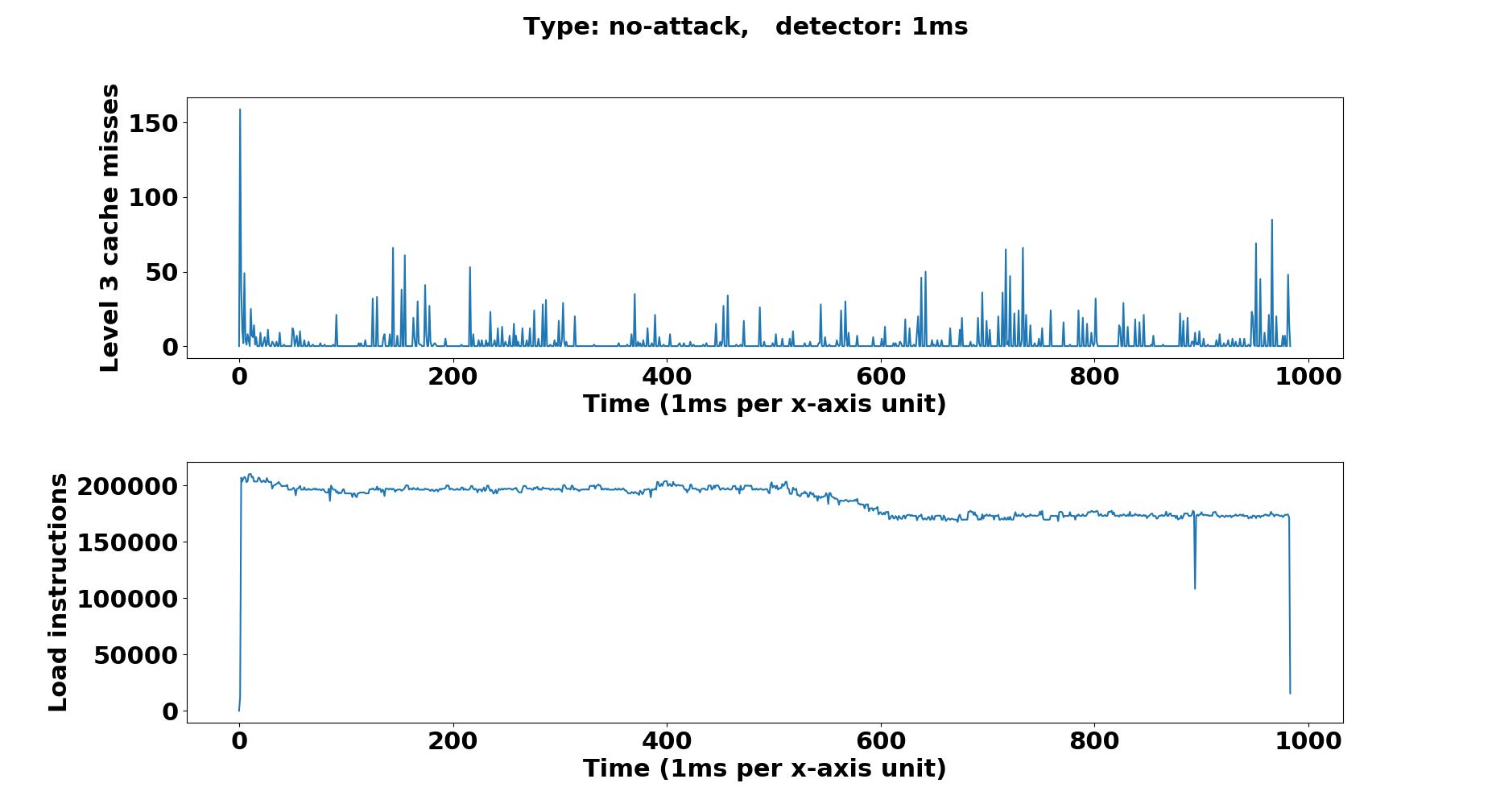}
        \caption{No attack. Detection freq.: 1 ms}
        \label{fig:subfig14}
\end{subfigure}

\begin{subfigure}[b]{0.49\textwidth}
        \includegraphics[width=\textwidth]{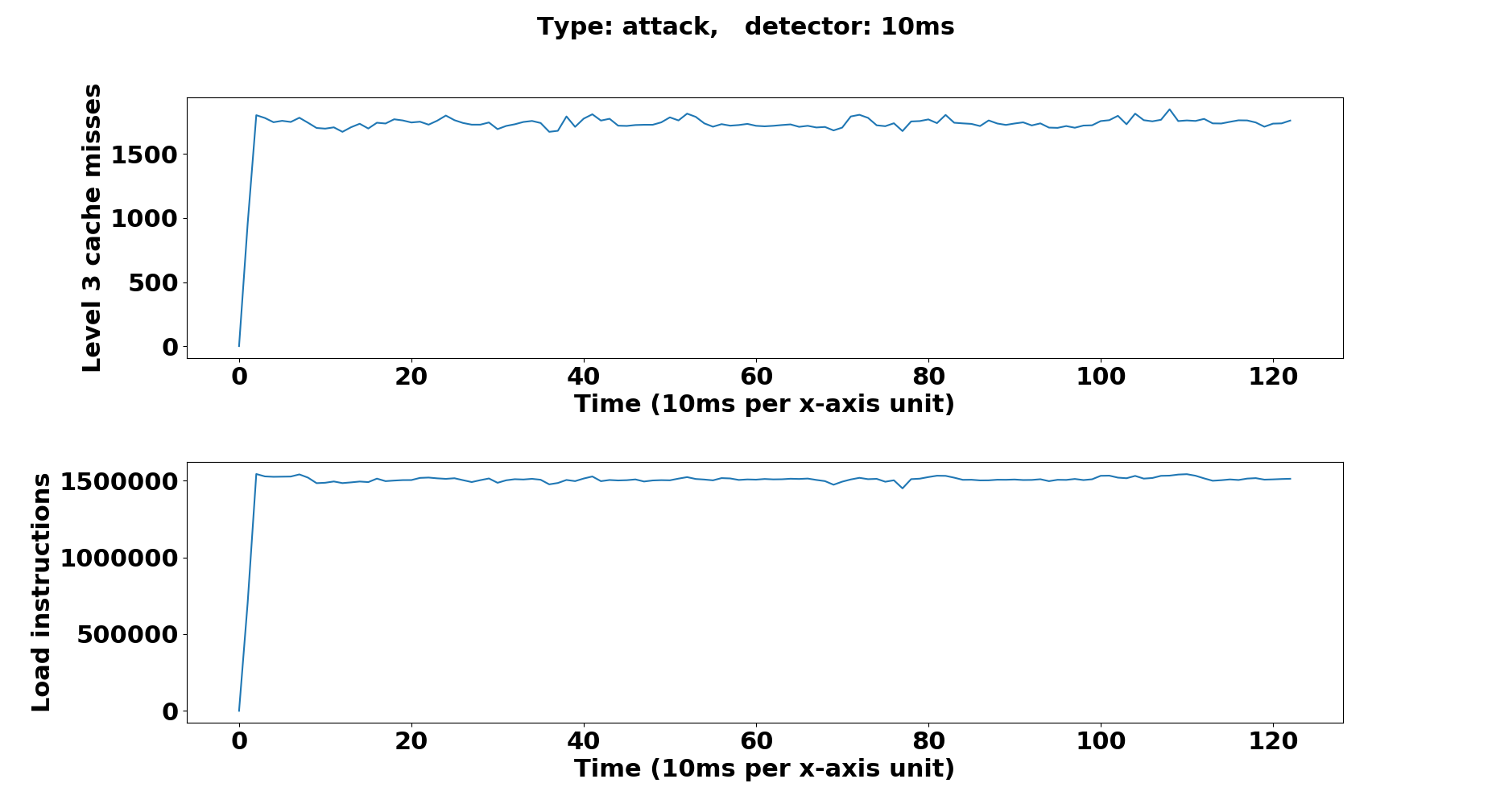}
        \caption{Attack. Detection freq.: 10 ms}
        \label{fig:subfig15}
\end{subfigure}
\begin{subfigure}[b]{0.49\textwidth}
        \includegraphics[width=\textwidth]{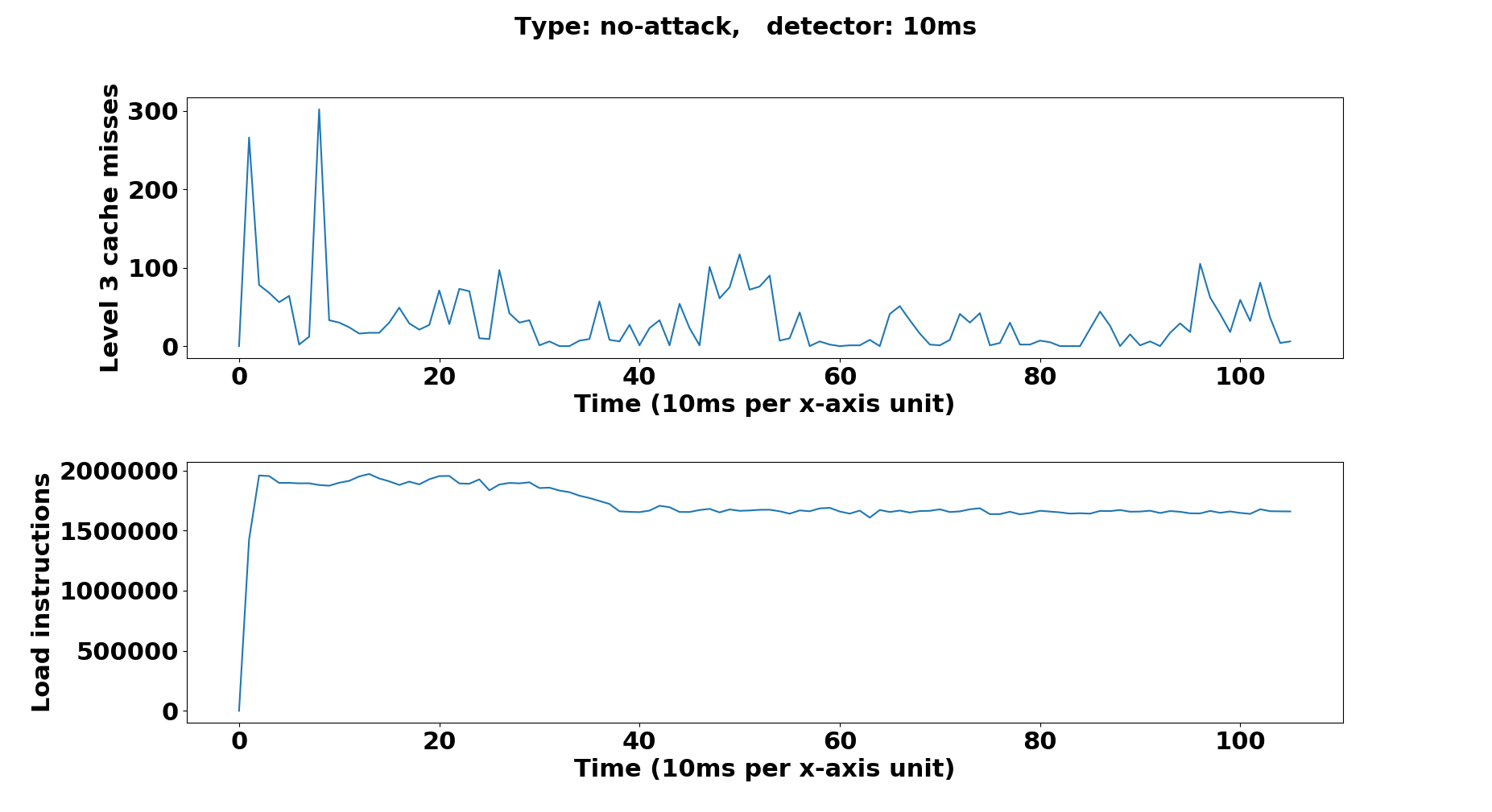}
        \caption{No attack. Detection freq.: 10 ms}
        \label{fig:subfig16}
\end{subfigure}

\caption{Results obtained from performance counters at different sampling rates. Each one of the three rows reports the results obtained for the L3 cache misses (above) and number of load instructions (below) in the victim under attack (left) and with no attack (right). The three rows correspond to the three sampling rates analyzed: 100 $\mu$s (first row), 1 ms (middle row) and 10 ms (last row).}
\label{fig:globfig1}
\end{figure}

\begin{figure}
\centering

\begin{subfigure}[b]{0.49\textwidth}
        \includegraphics[width=\textwidth]{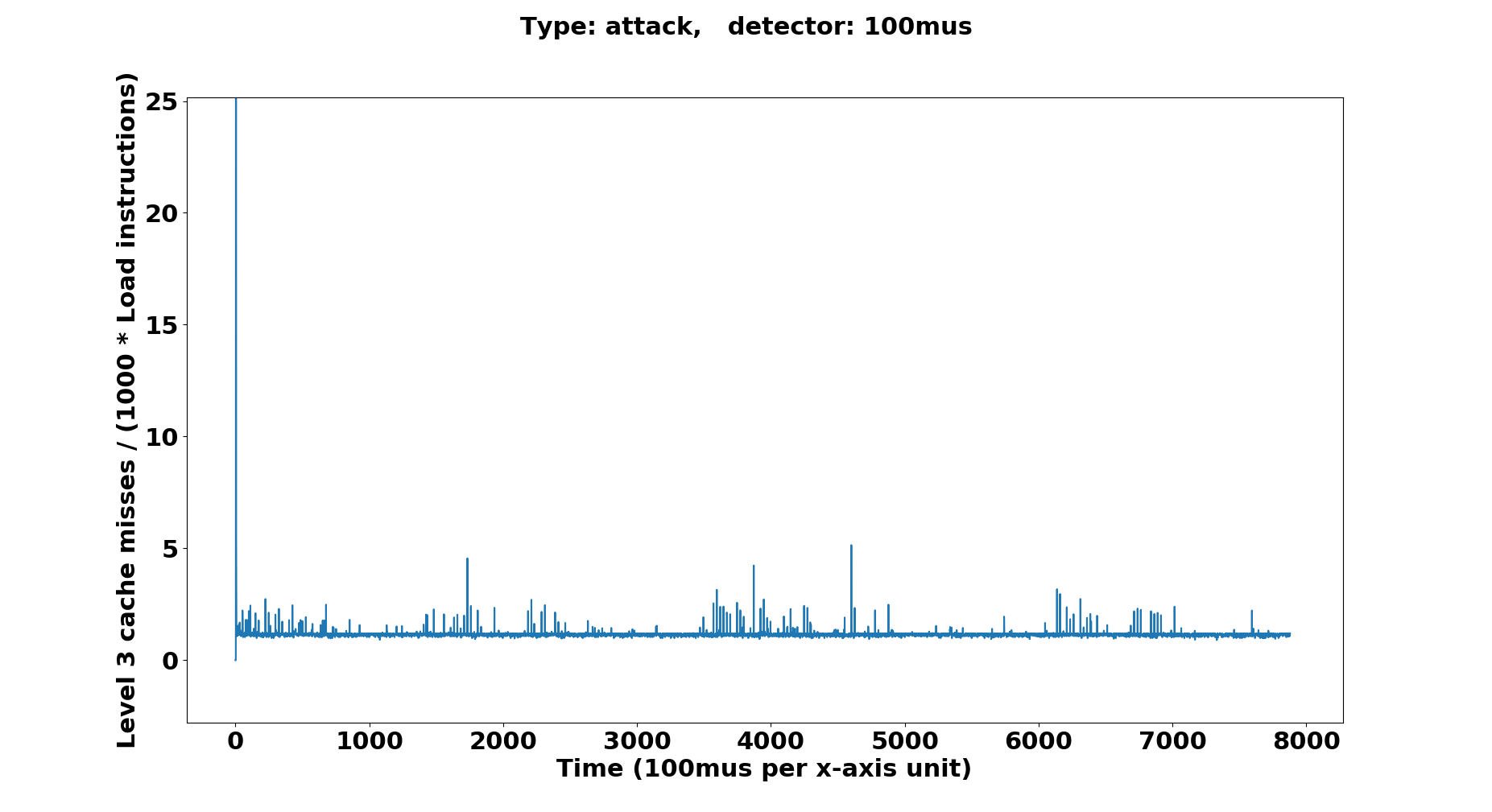}
        \caption{Attack. Detection freq.: 100 $\mu$s}
        \label{fig:subfig21}
\end{subfigure}
\begin{subfigure}[b]{0.49\textwidth}
        \includegraphics[width=\textwidth]{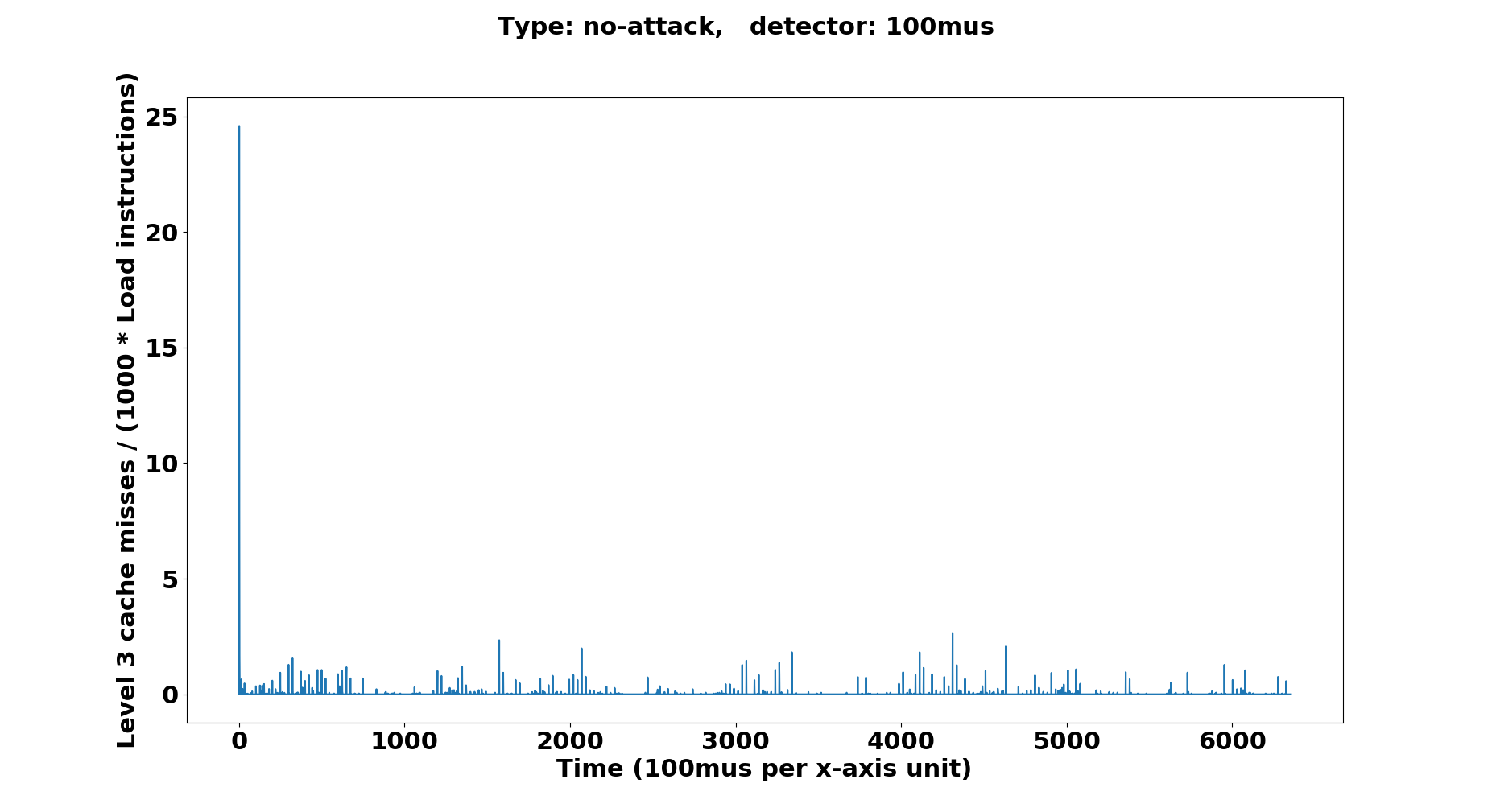}
        \caption{No attack. Detection freq.: 100 $\mu$s}
        \label{fig:subfig22}
\end{subfigure}

\begin{subfigure}[b]{0.49\textwidth}
        \includegraphics[width=\textwidth]{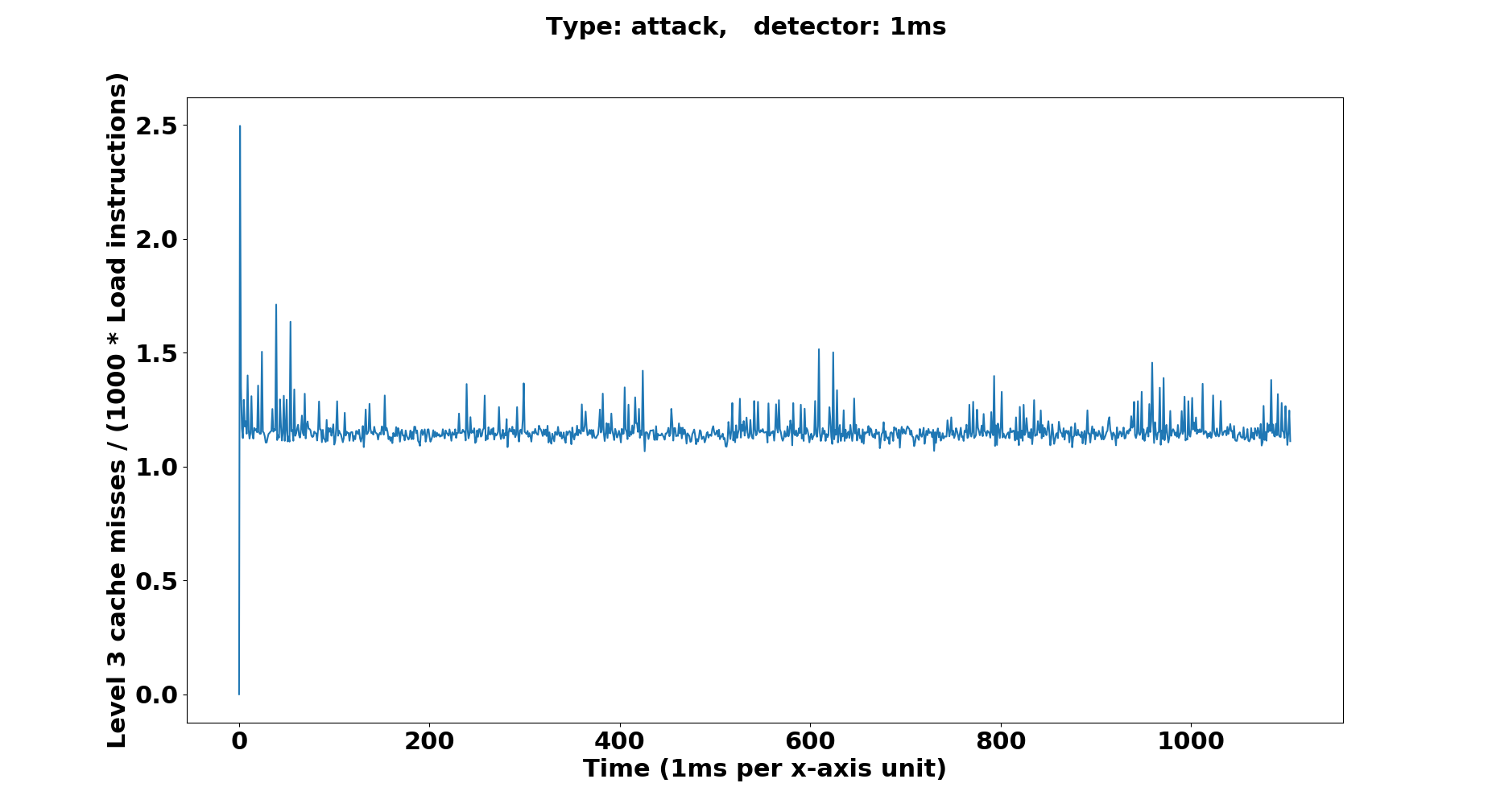}
        \caption{Attack. Detection freq.: 1 ms}
        \label{fig:subfig23}
\end{subfigure}
\begin{subfigure}[b]{0.49\textwidth}
        \includegraphics[width=\textwidth]{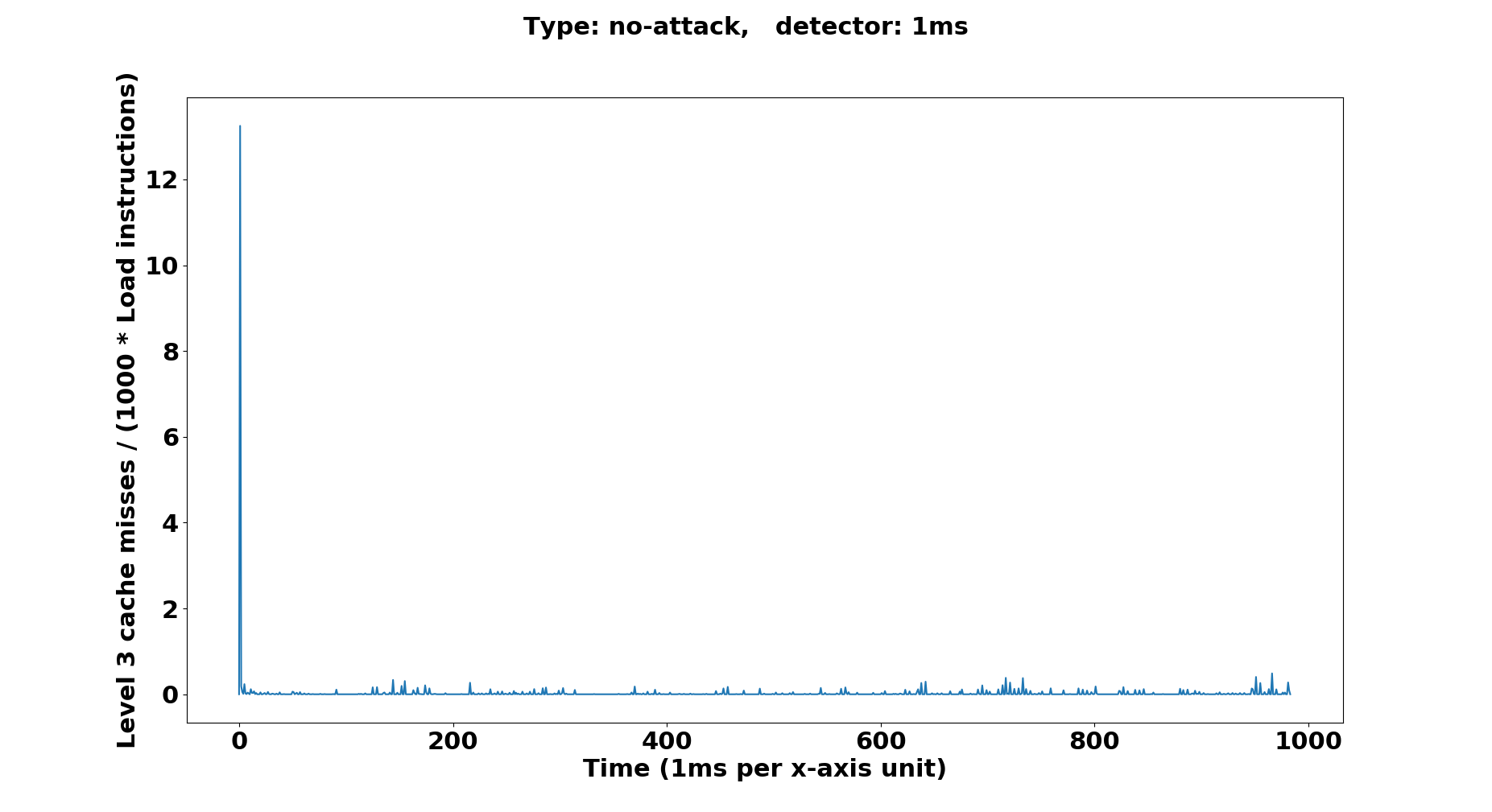}
        \caption{No attack. Detection freq.: 1 ms}
        \label{fig:subfig24}
\end{subfigure}

\begin{subfigure}[b]{0.49\textwidth}
        \includegraphics[width=\textwidth]{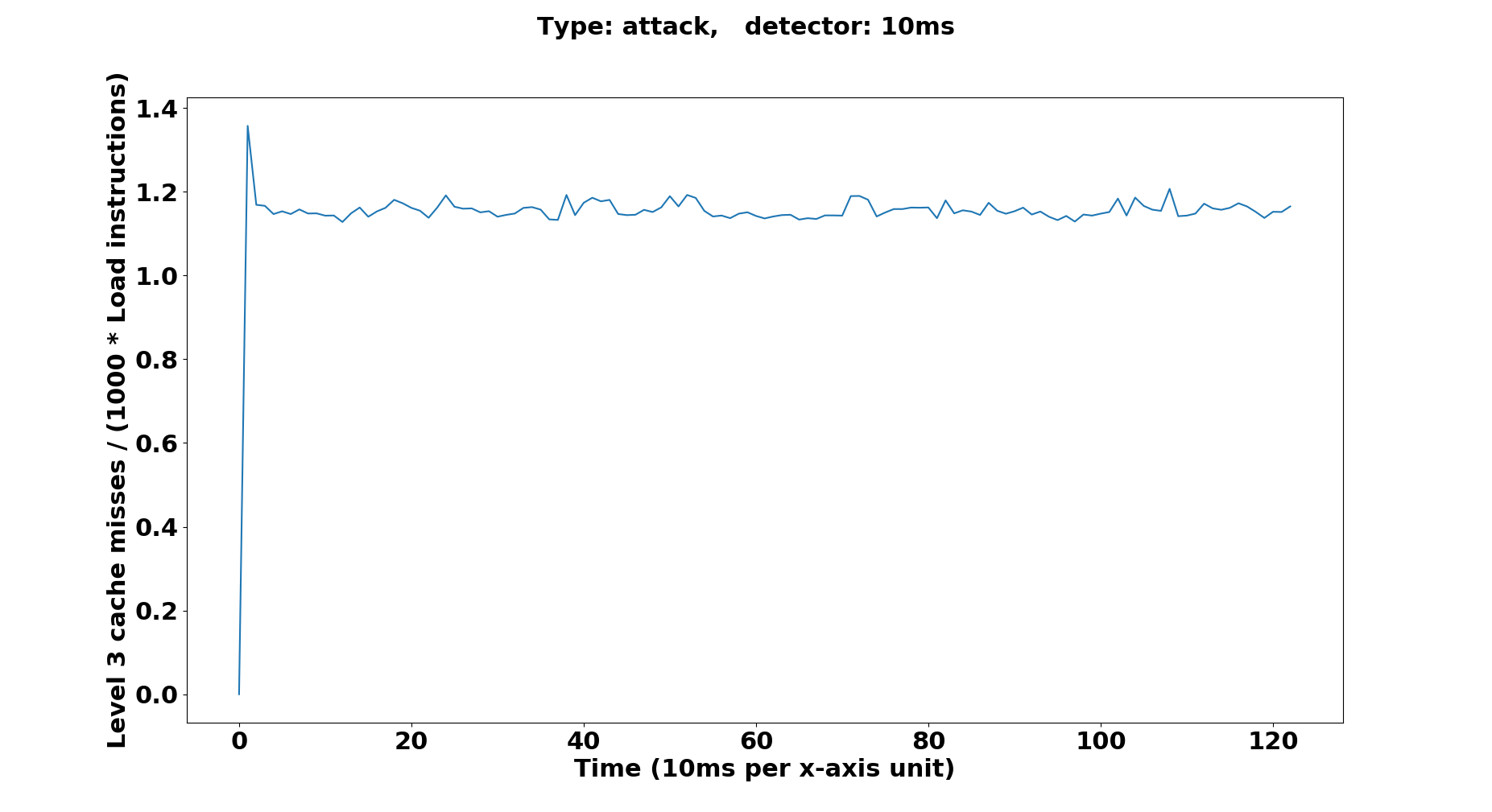}
        \caption{Attack. Detection freq.: 10 ms}
        \label{fig:subfig25}
\end{subfigure}
\begin{subfigure}[b]{0.49\textwidth}
        \includegraphics[width=\textwidth]{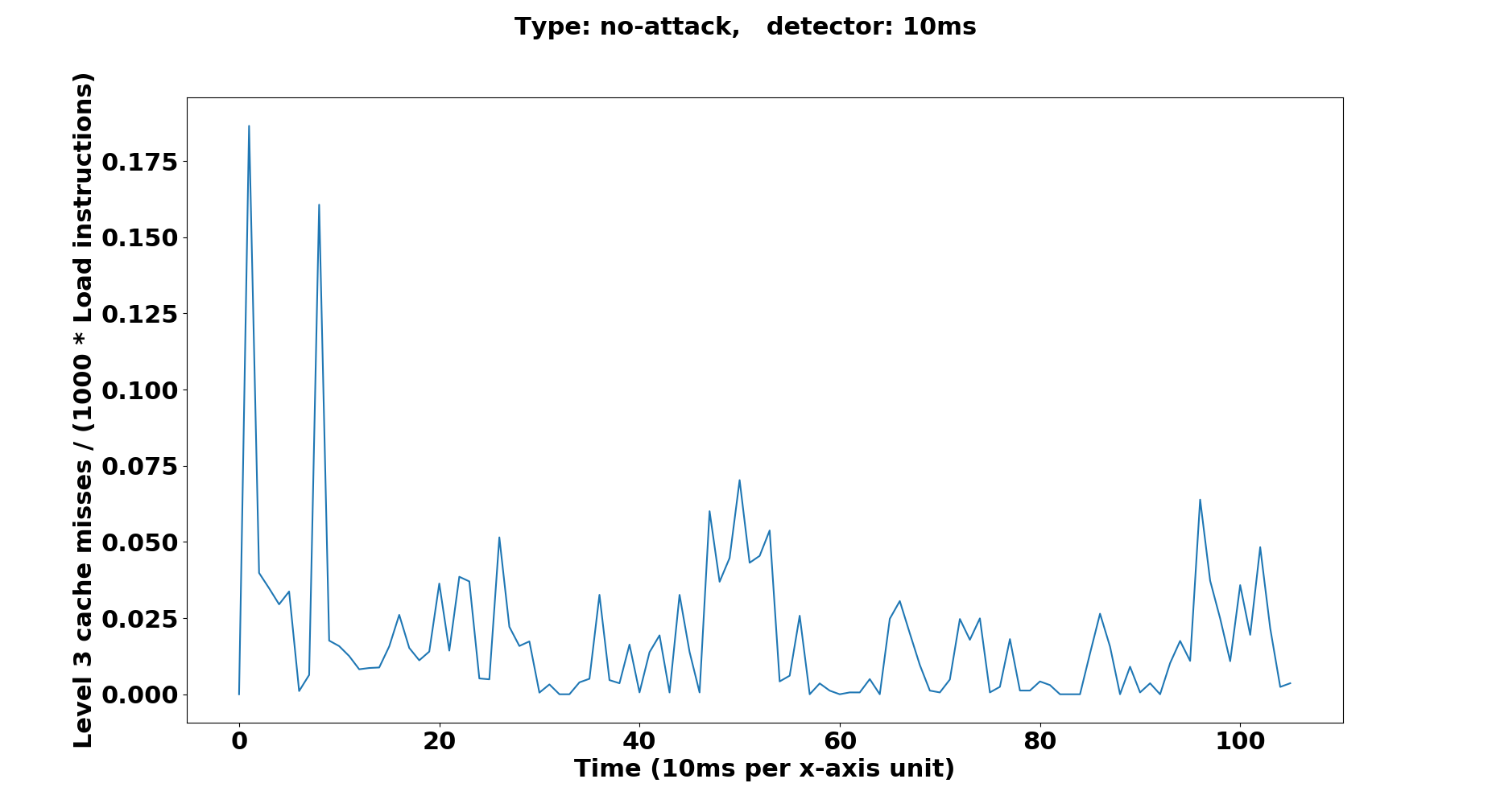}
        \caption{No attack. Detection freq.: 10 ms}
        \label{fig:subfig26}
\end{subfigure}

\caption{Metric evaluation for attack detection at different sampling rates. Each row displays the proposed metric: L3 cache misses per 1000 load instructions in the victim under attack (left-side column) and without attack (right-side column). The three rows correspond to the three sampling rates analyzed: 100 $\mu s$ (first row), 1 ms (middle row) and 10 ms (last row).}
\label{fig:globfig2}
\end{figure}

Figure~\ref{fig:globfig1} reports the observed PMC values chosen for the victim both in the presence and absence of attack. The experiment was repeated at different sampling frequencies, to study the effect of the sampling frequency in the detection capability.
Figure~\ref{fig:globfig2} shows the values of the proposed metric for the results in Figure~\ref{fig:globfig1}.
The first observation is that the selected metric is an effective mechanism to detect the attack; the values under attack are close to 1 while the values without attack are 10 to 100 times lower. In this situation,
the attack is detected if, after the initial cold misses (identified as 50 ms in our experiments),
the value remains close to 1.

A second conclusion from Figure~\ref{fig:globfig2} is that sampling PMCs at 100 $\mu s$ leads to more noisy 
results for the no-attack experiment. Given that this sampling rate also produces a higher overhead, 
we will not use that sampling frequency in the following.

\section{Analysis of a time-fragmented attack}
\label{sec:time-fragmented}

In this section, we propose a complete set of experiments in order to determine if the division of the
attack in discrete pieces and their distributed execution in time can potentially disguise the attack
and invalidate the action of our detector.

We proceed by dividing the $50,000$ encryptions needed for the attack into equally-sized groups
(or ``packets'' in the following) of encryptions. 
We have evaluated packets of decreasing sizes, namely: $5,000$, 500, 50 and 5 encryptions.
Furthermore, in order to analyze the effect of increasing the gap (time) between packets, 
we called ``interval'' the separation between two consecutive packets. 
In our experiments, we vary the interval between packets from 10 $\mu$s to 10 ms.
For each combination of packet size and interval we used the three sampling rates of the 
previous section: 100 $\mu$s, 1 ms, 10 ms.

\begin{figure}
\centering

\begin{subfigure}[b]{0.49\textwidth}
        \includegraphics[width=\textwidth]{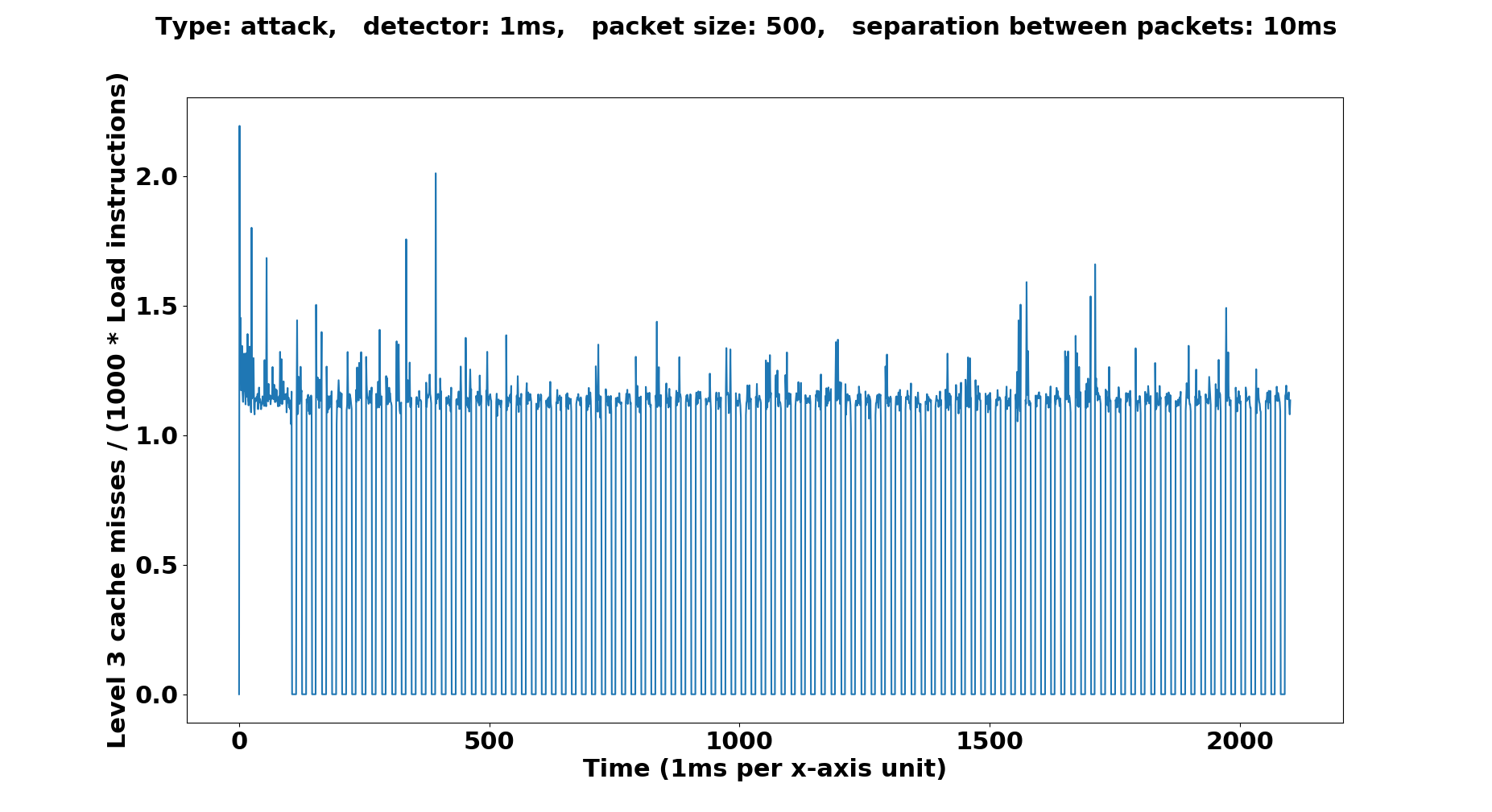}
        \caption{}
        \label{fig:subfig31}
\end{subfigure}
\begin{subfigure}[b]{0.49\textwidth}
        \includegraphics[width=\textwidth]{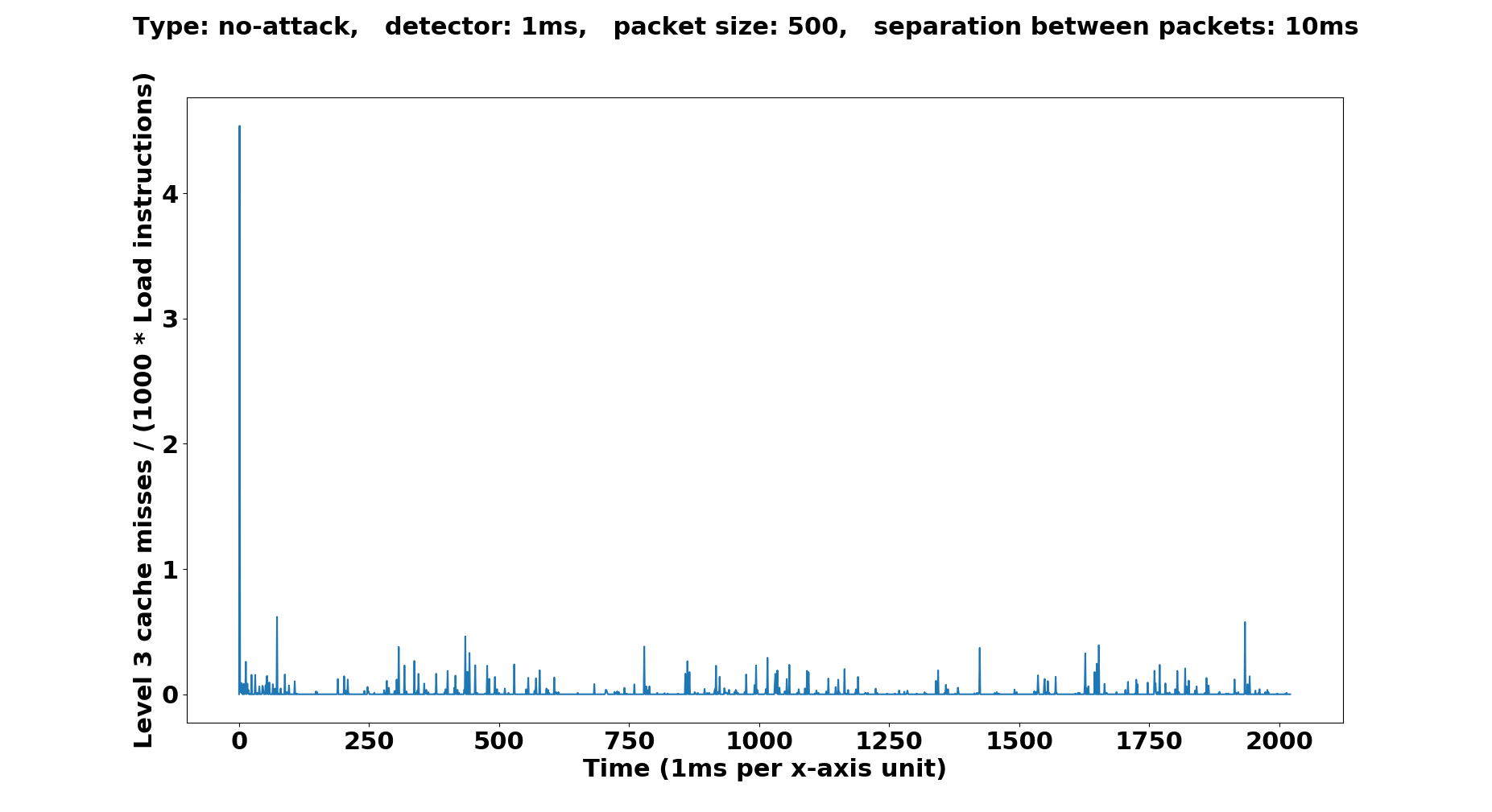}
        \caption{}
        \label{fig:subfig32}
\end{subfigure}

\begin{subfigure}[b]{0.49\textwidth}
        \includegraphics[width=\textwidth]{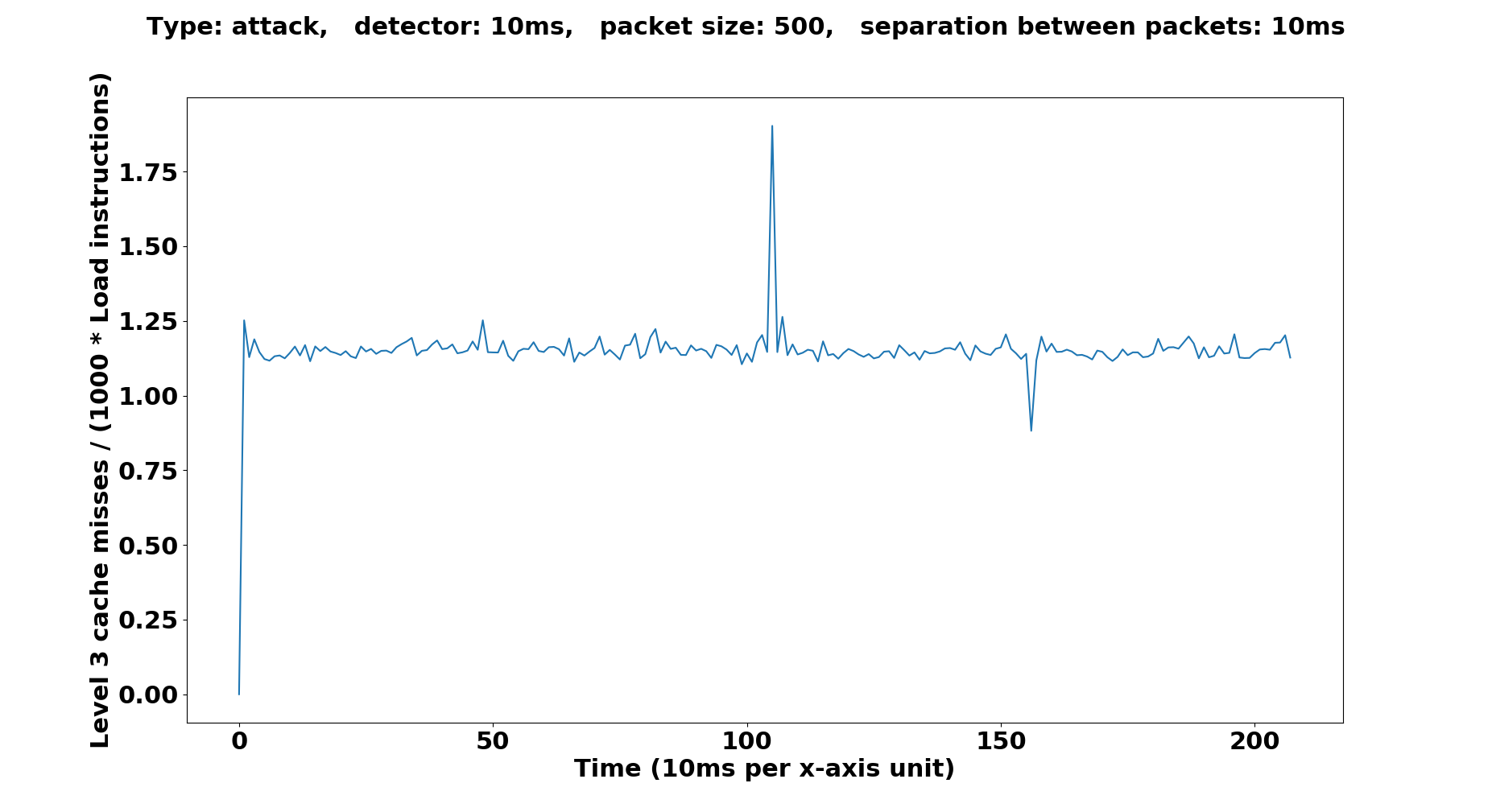}
        \caption{}
        \label{fig:subfig33}
\end{subfigure}
\begin{subfigure}[b]{0.49\textwidth}
        \includegraphics[width=\textwidth]{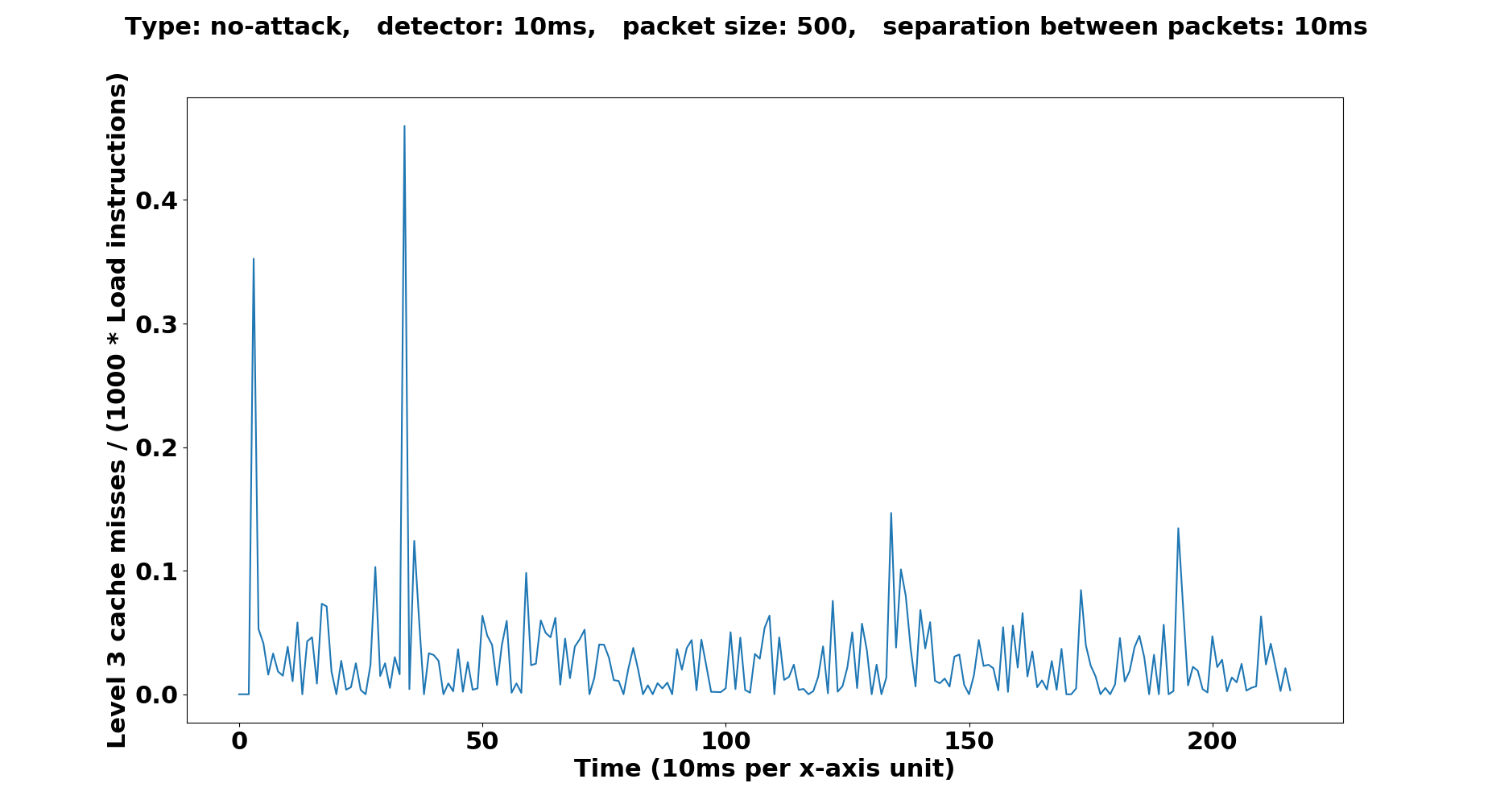}
        \caption{}
        \label{fig:subfig34}
\end{subfigure}

\caption{Detection metric for packets of 500 encryptions with interval of 10 ms. There is an attack in the left column and no attack in the right one. The sampling rate is 1 ms (top) and 10 ms (bottom).}
\label{fig:globfig3}

\end{figure}

The most interesting results are obtained for small packets and large intervals, as expected.
Figure \ref{fig:globfig3} shows the results when the attack is divided into 10 packets of 500 encryptions, 
and the time interval between two consecutive packets is 10 ms. The sampling rate is either 1ms or 10 ms. The metrics obtained from the 10 ms samples are close to the usual value for the attack, but the results for 1 ms samples switch from the values corresponding to an attack (close to 1) to the no-attack values (close to 0).
As expected, for the high resolution frequency some samples do not find any difference between attack and no-attack, because they fall in the interval of time between packets of the attack. On the contrary, the low resolution samples always find the ``big picture''.

\begin{figure}
\centering

\begin{subfigure}[b]{0.49\textwidth}
        \includegraphics[width=\textwidth]{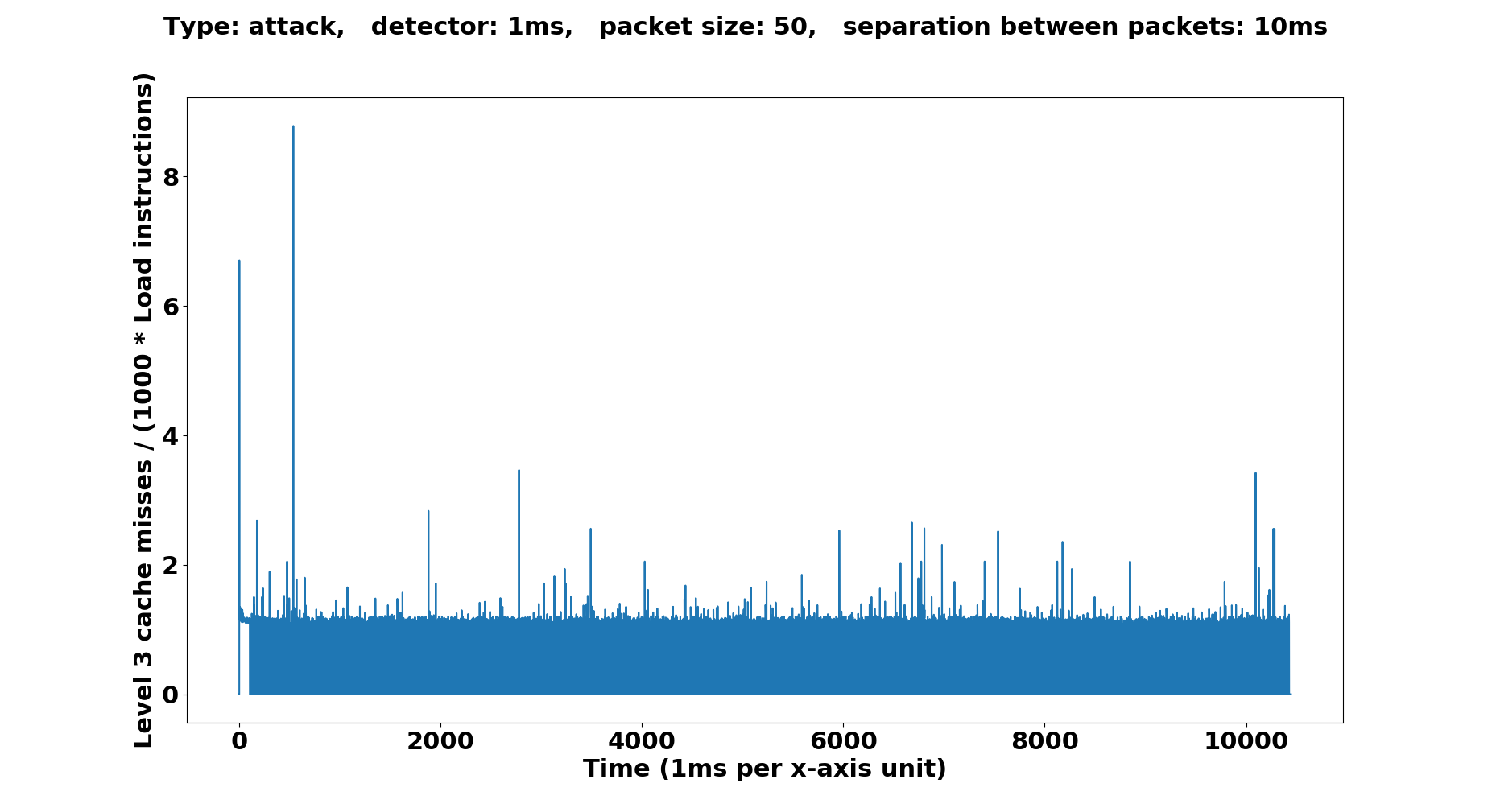}
        \caption{}
        \label{fig:subfig41}
\end{subfigure}
\begin{subfigure}[b]{0.49\textwidth}
        \includegraphics[width=\textwidth]{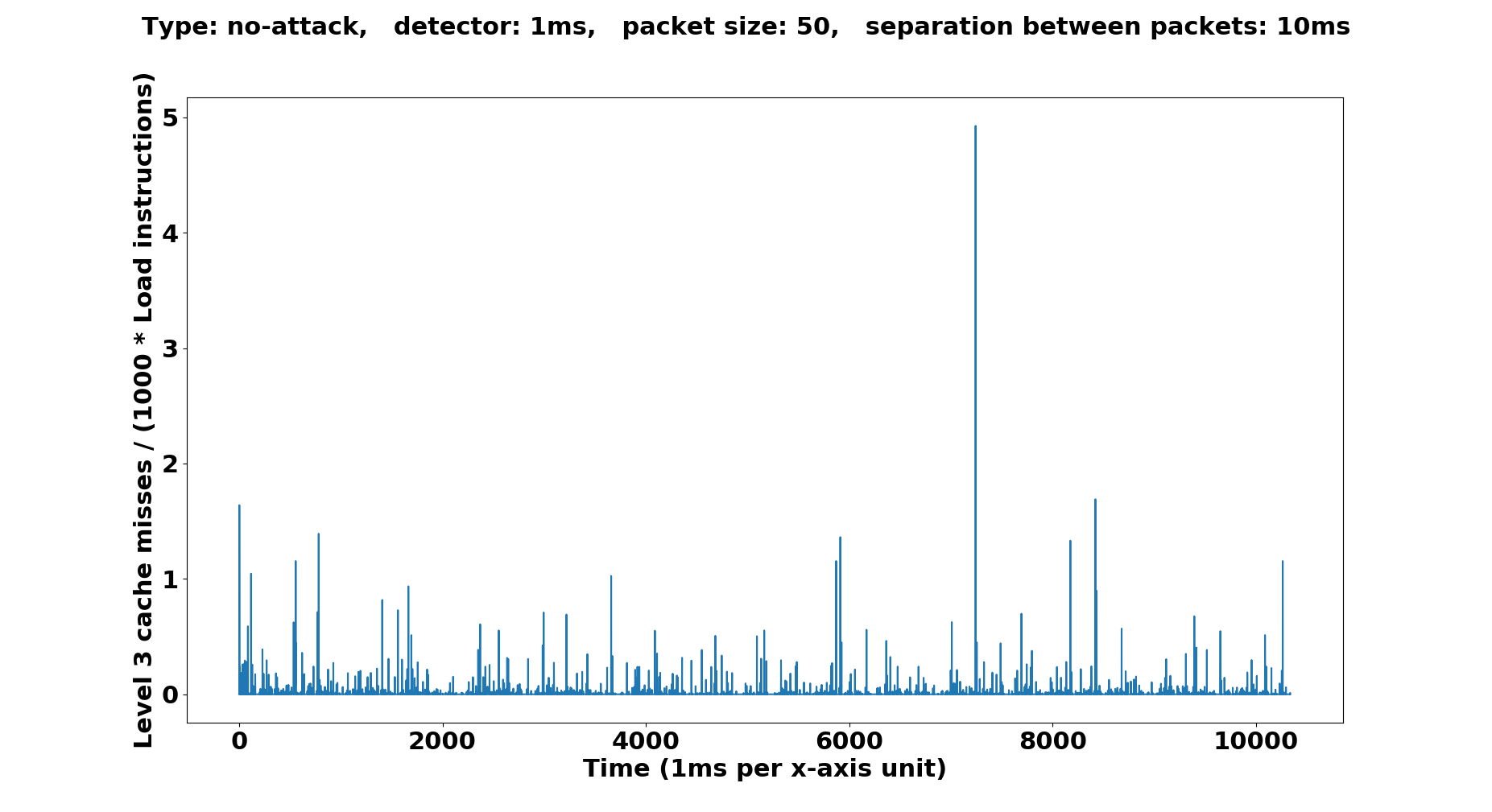}
        \caption{}
        \label{fig:subfig42}
\end{subfigure}

\begin{subfigure}[b]{0.49\textwidth}
        \includegraphics[width=\textwidth]{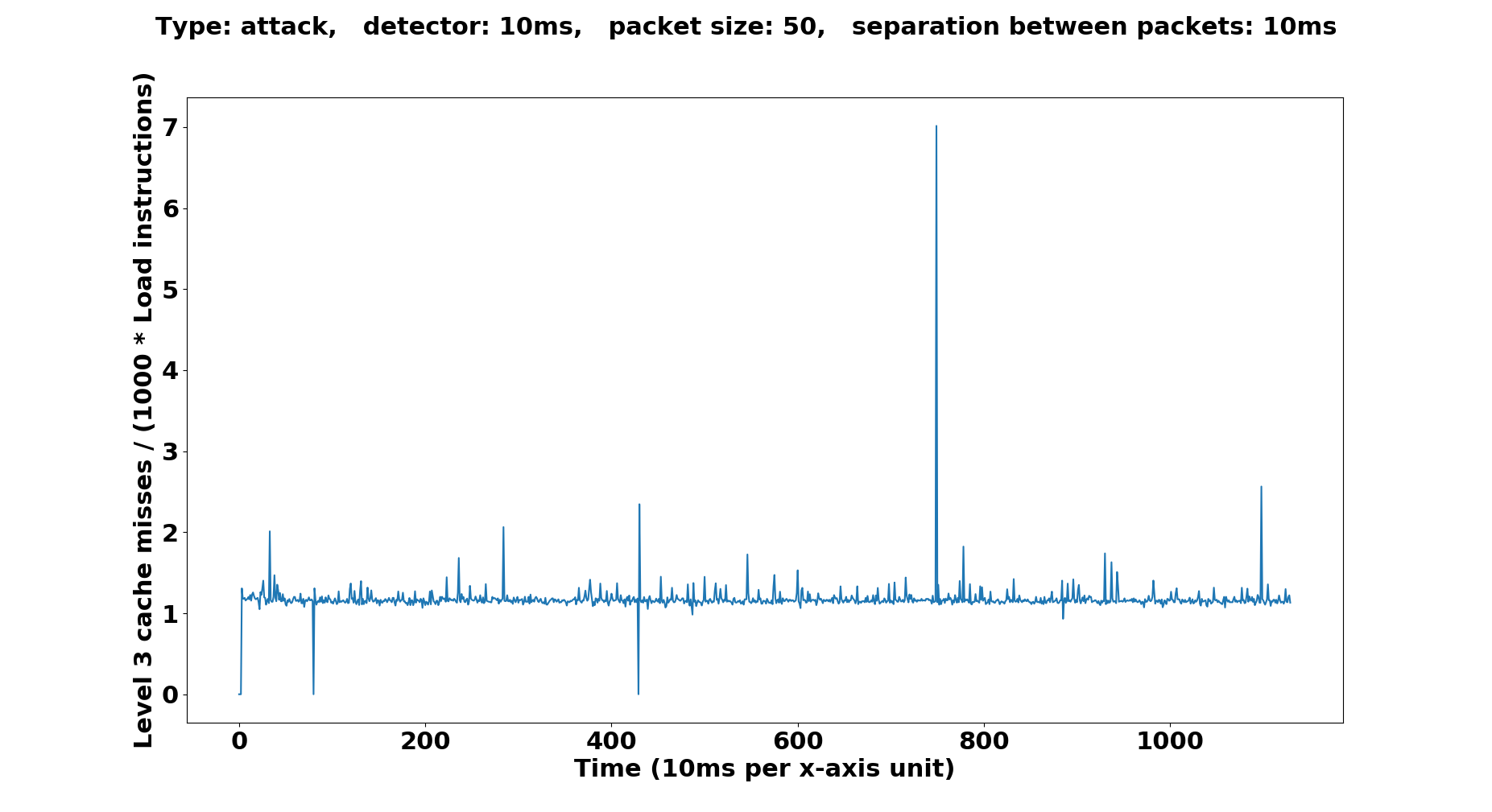}
        \caption{}
        \label{fig:subfig43}
\end{subfigure}
\begin{subfigure}[b]{0.49\textwidth}
        \includegraphics[width=\textwidth]{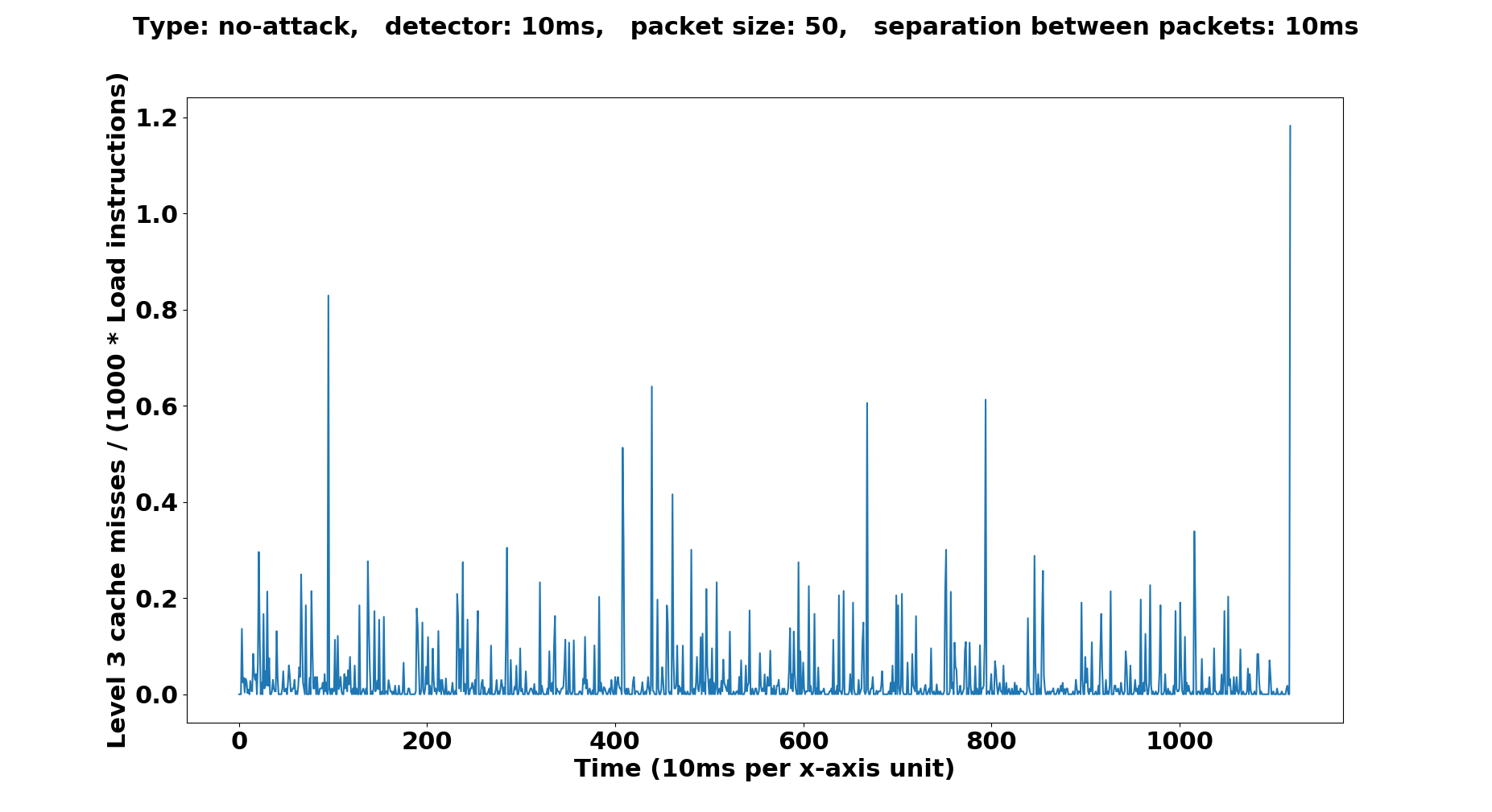}
        \caption{}
        \label{fig:subfig44}
\end{subfigure}

\caption{Detection metric for packets of 50 encryptions with interval of 10 ms. There is an attack in the left column and no attack in the right one. The sampling rate is 1 ms (top) and 10 ms (bottom).}
\label{fig:globfig4}
\end{figure}

\begin{figure}
    \centering
    \includegraphics[width=0.8\textwidth]{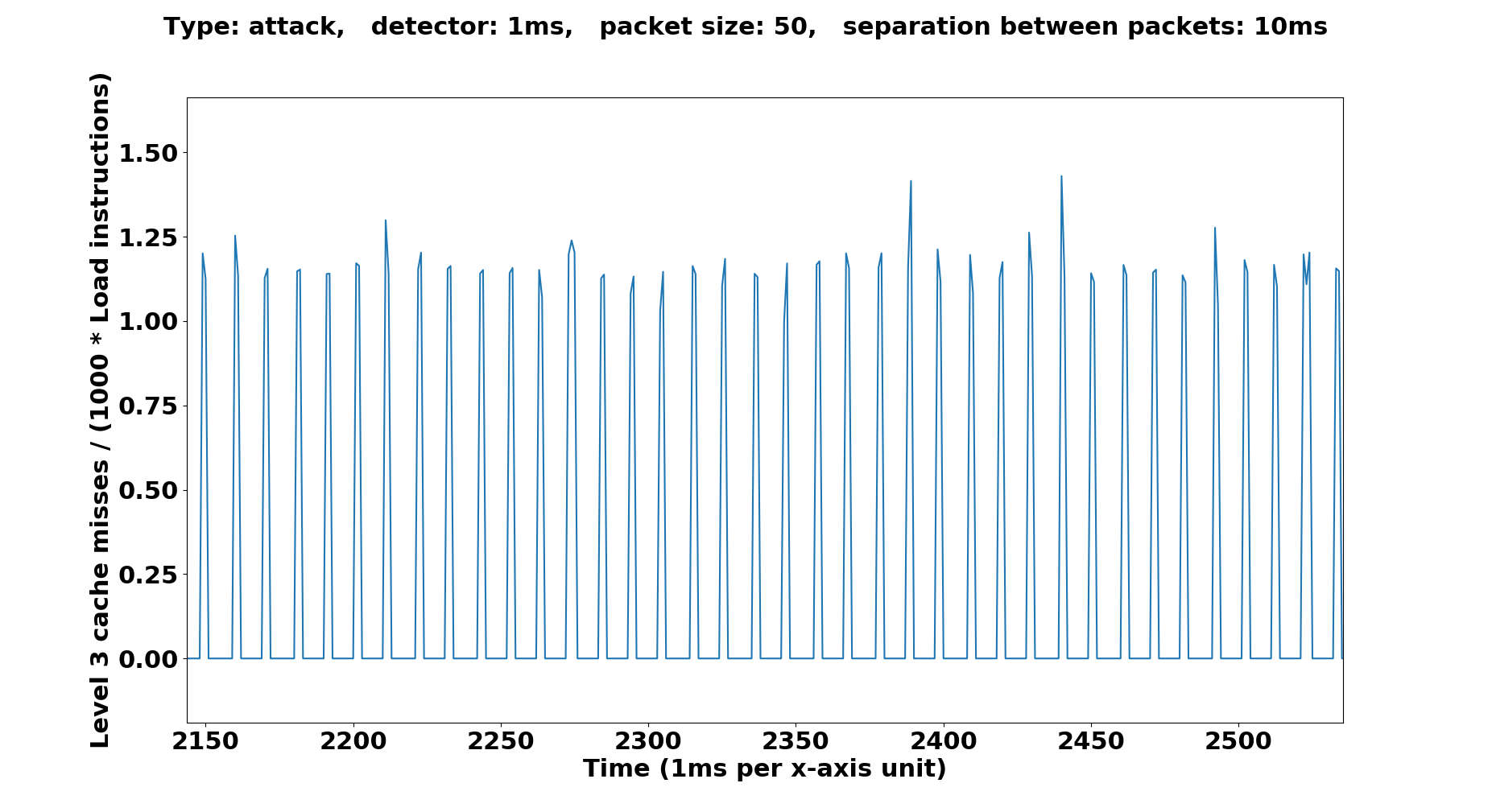}
    \caption{Augmented view of the detection metric for the attack with packets of 50 encryptions, interval of 10 ms and sampling rate of 1 ms.}
    \label{fig:detailp5-1ms}
\end{figure}

Figure~\ref{fig:globfig4} reports an equivalent evaluation for packets 10 times smaller, with
the aim of reducing the time in which the attack can be detected. In this case, the difference between the obtained results at different sampling rates is more evident. For the 1 ms sampling rate, on one hand, the no-attack experiment has higher number of L3 misses due to the separation between packets. During those intervals some cache lines are evicted due to normal functioning of the system. On the other hand, the experiment with attack also switches from low to high values of the metric as in the previous experiment. This fact can be observed in Figure~\ref{fig:detailp5-1ms}, which is an augmented view of the results for the attack with 1 ms sampling. It confirms that the attack can be more easily hidden from the high resolution samples than from the lower ones.

\begin{figure}
\centering

\begin{subfigure}[b]{0.49\textwidth}
        \includegraphics[width=\textwidth]{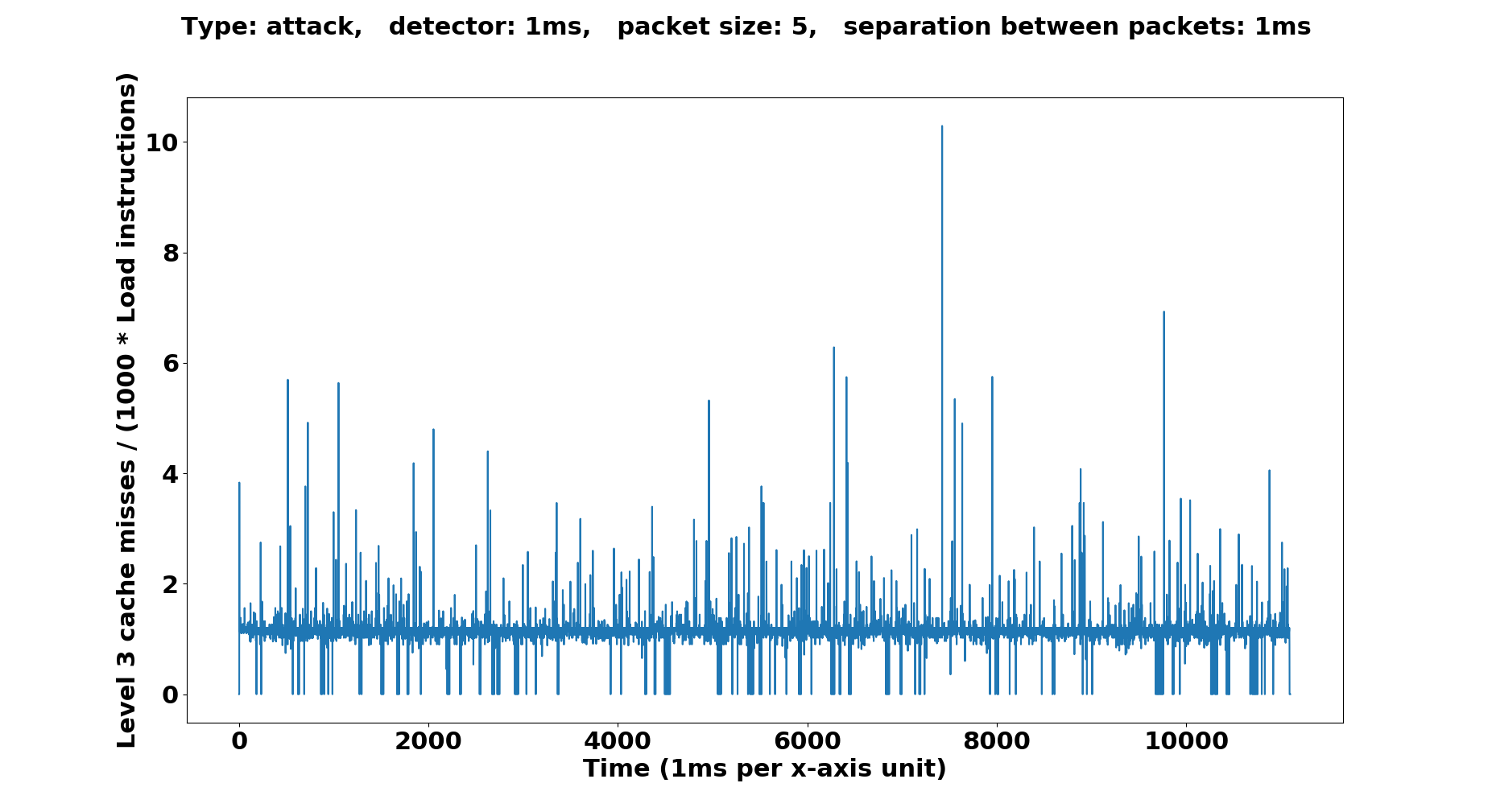}
        \caption{}
        \label{fig:subfig41}
\end{subfigure}
\begin{subfigure}[b]{0.49\textwidth}
        \includegraphics[width=\textwidth]{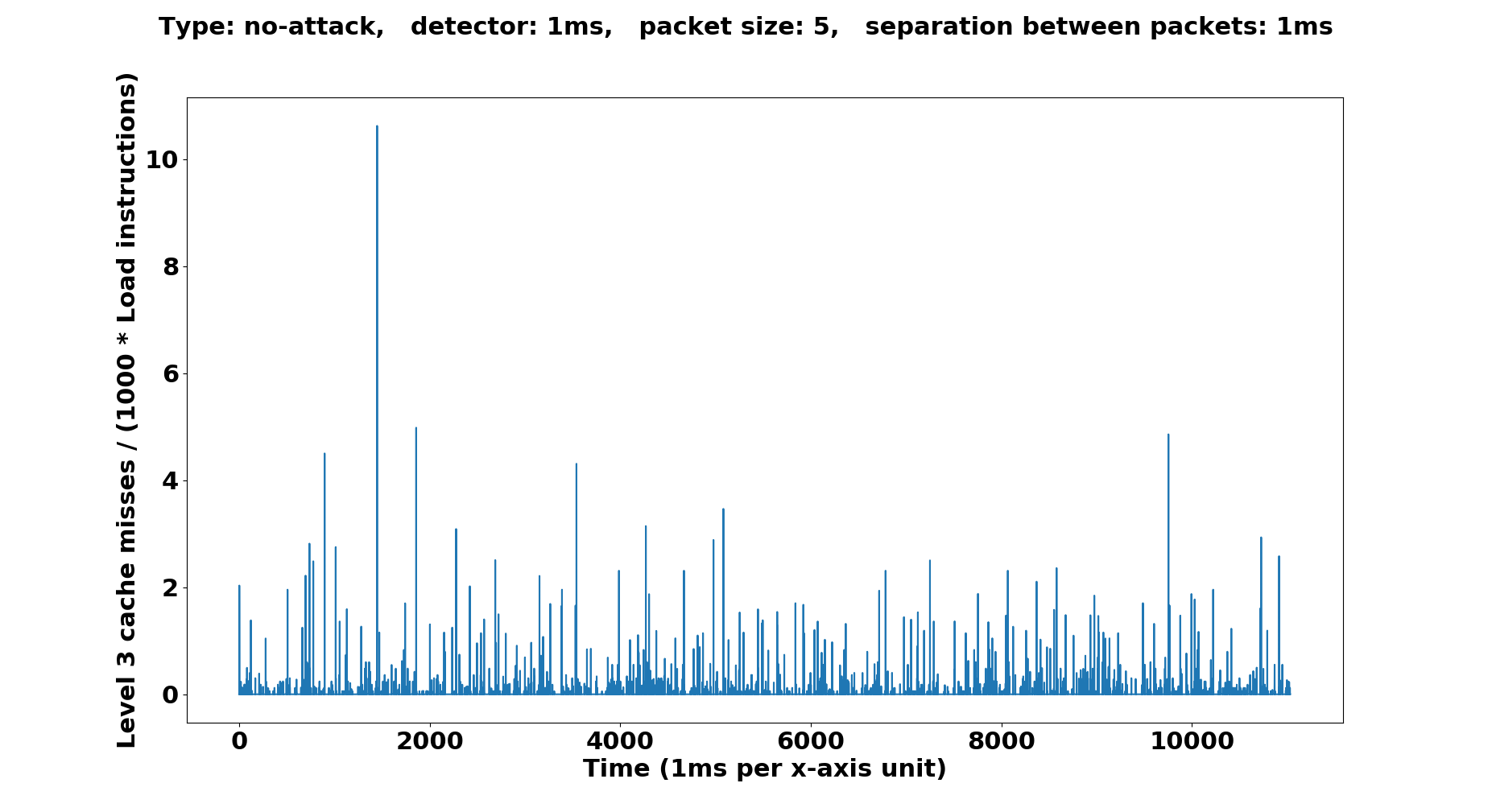}
        \caption{}
        \label{fig:subfig42}
\end{subfigure}

\begin{subfigure}[b]{0.49\textwidth}
        \includegraphics[width=\textwidth]{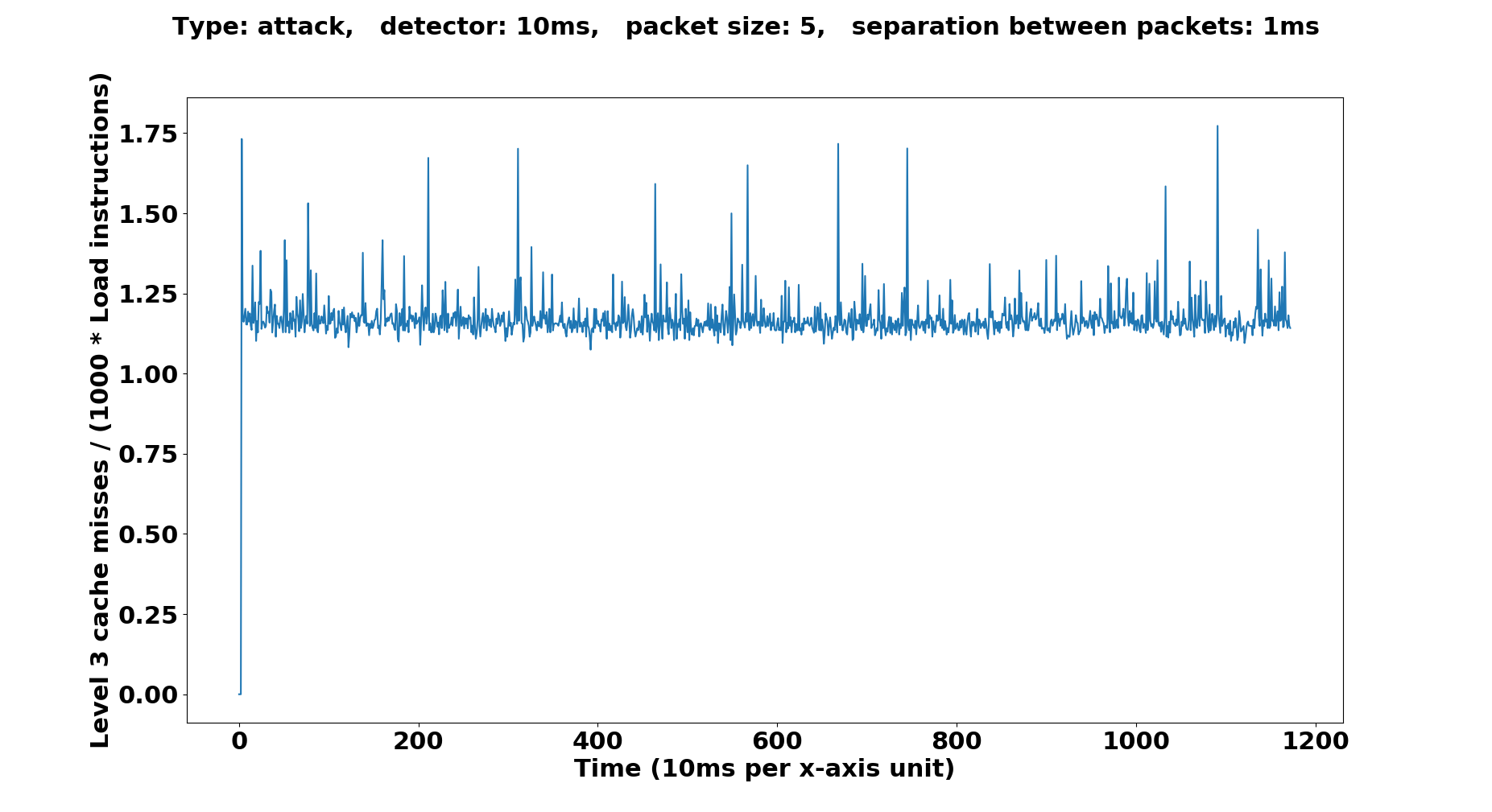}
        \caption{}
        \label{fig:subfig43}
\end{subfigure}
\begin{subfigure}[b]{0.49\textwidth}
        \includegraphics[width=\textwidth]{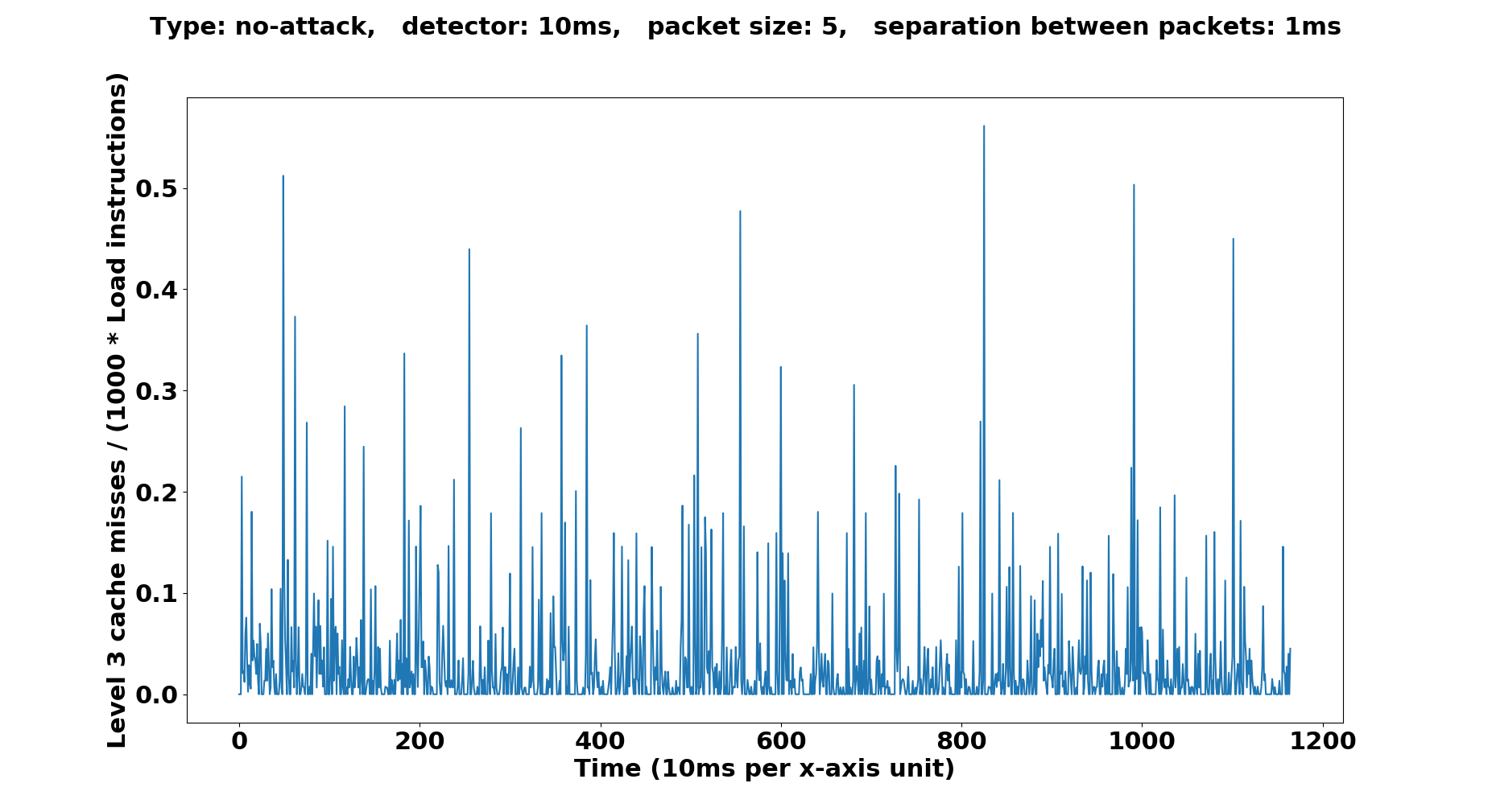}
        \caption{}
        \label{fig:subfig44}
\end{subfigure}

\caption{Detection metric for packets of 5 encryptions with interval of 1 ms. There is an attack in the left column and no attack in the right one. The sampling rate is 1 ms (above) and 10 ms (below).}
\label{fig:globfig5}
\end{figure}

Finally, the packet size is decreased to 5 encryptions. In this experiment, when the interval between packets is longer than 1 ms the attack stops working. The results for 1 ms interval are show in 
Figure~\ref{fig:globfig5} and they confirm that our detection metric is able to detect the attack with a sampling rate of 10 ms.

\section{Conclusions}
\label{sec:conclusions}
In this paper, we proposed a mechanism to protect victim processes running in multi-core servers (either native or inside a VM) against cache timing attacks by adding to the server a new detector process that monitors only the PMCs associated to the victim process.
To that end, we implemented a cache timing attack against the table based AES encryption algorithm. We used L3 cache misses per 1000 load instructions as a detection metric and achieved detection of the attack for all the different sampling rates, although sampling at high frequency is worse than at lower ones.

We have tried to hide the attack dividing it into small parts and interleaving time slots with attack and without attack. Thus, sampling PMC at high frequency makes detection of the attack more difficult. Again, lower frequency monitoring results in higher detection capability. 

\section{Acknowledgements}
This work is supported by the EU FEDER and the Spanish MINECO under grant number TIN2015-65277-R and by Spanish CM under grant S2018/TCS-4423. We would like to thank Samira Briongos and Pedro Malag\'{o}n for their helpful comments on some details of the attack implementation.
%
%

\bibliographystyle{./styles/bibtex/spmpsci.bst}
\bibliography{./references.bib}

\end{document}